\documentclass[a4paper,fleqn,usenatbib]{mnras}

\usepackage{newtxtext,newtxmath}

\usepackage[T1]{fontenc}
\usepackage{ae,aecompl}

\usepackage{graphicx}	
\usepackage{amsmath}	
\usepackage{amssymb}	
\usepackage{multicol}        
\usepackage{bm}		
\usepackage{pdflscape}	
\usepackage{csquotes}
\usepackage{hyperref}

\title[Hunting for Metals in XQ-100 Composite Spectrum]{Hunting for Metals Using XQ-100 Legacy Survey Composite Spectra}

\author[S. Perrotta, V. D'Odorico, F. Hamann et al.]{
S. Perrotta,$^{1,2}$\thanks{E-mail: serenap@ucr.edu (SP)}
V. D'Odorico,$^{2,3}$
F. Hamann,$^{1}$
S. Cristiani,$^{2,4}$
J. X. Prochaska,$^{5}$
\newauthor
S. L. Ellison,$^{6}$
S. Lopez,$^{7}$
G. Cupani,$^{2}$
G. Becker,$^{1}$
T.  A. M. Berg,$^{6}$
L. Christensen,$^{8}$
\newauthor
K. D. Denney$^{9}$
and G. Worseck$^{10}$
\\
$^{1}$Department of Physics and Astronomy, University of California, 900 University Avenue, Riverside, CA 92521, USA\\
$^{2}$INAF-OATS, Via Tiepolo 11, 34143 Trieste, Italy\\
$^{3}$Scuola Normale Superiore, Palazzo D'Ancona, Via Consoli del Mare, 1, 56126 Pisa PI\\
$^{4}$INFN-National Institute for Nuclear Physics, via Valerio 2, I-34127 Trieste, Italy\\
$^{5}$Astronomy \& Astrophysics, UC Santa Cruz, 1156 High St., Santa Cruz, CA 95064 USA\\
$^{6}$Department of Physics \& Astronomy, University of Victoria, Finnerty Road, Victoria, British Columbia, V8P 1A1, Canada\\
$^{7}$Departamento de Astronomía, Universidad de Chile, Casilla 36-D, Santiago\\
$^{8}$Dark Cosmology Centre, Niels Bohr Institute, University of Copenhagen, Juliane Maries Vej 30, DK-2100 Copenhagen, Denmark\\
$^{9}$Department of Astronomy, The Ohio State University, 140 West 18th Avenue, Columbus, OH 43210, USA\\
$^{10}$Institut f\"{u}r Physik und Astronomie, Universit\"{a}t Potsdam, Karl-Liebknecht-Str. 24/25, D-14476 Potsdam, Germany
}

\date{Accepted 2018 August 10. Received 2018 August 9; in original form 2018 May 29}

\pubyear{2018}

\begin{document}
\label{firstpage}
\pagerange{\pageref{firstpage}--\pageref{lastpage}}
\maketitle

\begin{abstract}

We investigate the N{\sevensize V} absorption signal along the line of sight of background quasars, in order to test the robustness of the use of this ion as criterion to select intrinsic (i.e. physically related to the quasar host galaxy) narrow absorption lines (NALs).
We build composite spectra from a sample of $\sim$ 1000 C{\sevensize IV} absorbers, covering the redshift range 2.55 < z < 4.73, identified in 100 individual sight lines from the XQ-100 Legacy Survey.
We detect a statistical significant N{\sevensize V} absorption signal only within 5000 km s$^{-1}$ of the systemic redshift, z$\rm_{em}$. This absorption trough is $\sim$ 15$\sigma$ when only C{\sevensize IV} systems with N(C{\sevensize IV}) > 10$^{14}$ cm$^{-2}$ are included in the composite spectrum. This result confirms that N{\sevensize V} offers an excellent statistical tool to identify intrinsic systems. We exploit the stacks of 11 different ions to show that the gas in proximity to a quasar exhibits a considerably different ionization state with respect to gas in the transverse direction and intervening gas at large velocity separations from the continuum source. Indeed, we find a dearth of cool gas, as traced by low-ionization species and in particular by Mg{\sevensize II}, in the proximity of the quasar. 
We compare our findings with the predictions given by a range of Cloudy ionization models and find that they can be naturally explained by ionization effects of the quasar.

\end{abstract}

\begin{keywords}
(galaxies:) quasars: absorption lines -- (galaxies:) intergalactic medium -- galaxies: high-redshift
\end{keywords}



\section{INTRODUCTION}

Quasar outflows have been increasingly invoked by popular evolution models to regulate both star formation in host galaxies and the accretion of material on to central supermassive black holes (SMBH; e.g. \citealp{Granato2004, DiMatteo2005, Hopkins2016, Weinberger2017}). A SMBH at the centre of a galaxy can produce a large amount of energy ($\sim$  10$^{62}$ erg). Even if just a few per cent of the quasar bolometric luminosity were injected into the interstellar medium (ISM) of the host galaxy, it could have a significant impact on the host galaxy evolution \citep{Scannapieco2004, Prochaska2009, Hopkins2010}. 
Such feedback offers a natural explanation for the observed mass correlation between SMBHs and their host galaxy spheroids (e.g., \citealp{King2003, McConnell2013}). However, the actual mechanisms of feedback remain highly uncertain.

Outflows are often studied in the rest-frame ultraviolet (UV) via blueshifted narrow absorption lines (NALs) defined by full width at half maximum (FWHM) $\lesssim$ 300 km s$^{-1}$.
These absorbers are ubiquitously found and detected in all active galactic nuclei (AGN) subclasses (e.g., \citealp{Crenshaw1999, Ganguly2001, Vestergaard2003, Misawa2007}). 
The limitation of NALs is that they arise from a wide range of environments, from high-speed outflows to halo gas to physically unrelated gas or galaxies at large distances from the quasar. Those forming in the proximity of quasars provide valuable tools to study the gaseous environment of quasar host galaxies (e.g. \citealp{Dodorico2004, Wu2010, Berg2018}).

With the aim of investigating quasar winds, we must select NALs that truly trace outflowing gas cautiously. The outflow/intrinsic origin of individual NALs can be inferred from: i) time variability of line profiles, ii) absorption profiles significantly broader and smoother compared with the thermal line widths, iii) high space densities measured directly from excited-state fine-structure lines, iv) partial coverage of the background emission source measured via resolved, optically thick lines with too-shallow absorption troughs, v) higher ionization states than intervening absorbers. Nonetheless, NALs can still be connected to the quasar host galaxy without exhibiting such properties \citep{Hamann1997}.

Previous works have shown that, among the highly-ionized absorption species commonly used to identify outflows (e.g. C{\sevensize IV}$\lambda \lambda$1548,1550; Si{\sevensize IV}$\lambda \lambda$1393,1402; N{\sevensize V}$\lambda \lambda$1238,1242; O{\sevensize VI}$\lambda \lambda$1031,1037), N{\sevensize V} is detected with the lowest frequency, but with the highest intrinsic fraction.
\citet{Misawa2007} derived an intrinsic fraction of 75 per cent for N{\sevensize V} NALs, through partial coverage and line locking methods. \citet{Ganguly2013}, using the same techniques, found a value of 29-56 per cent and suggested using this ion in building large catalogs of intrinsic NALs with lower resolution and/or lower signal-to-noise ratio (S/N) data. 

In \citeauthor{Perrotta2016} (\citeyear{Perrotta2016}, hereafter P16) we used the spectra of 100 quasars at emission redshift z$\rm _{em}$ = 3.5 - 4.72 to build a large, relatively unbiased, sample of NALs and study their physical properties statistically. The spectra have been obtained with the echelle spectrograph X-shooter \citep{Vernet2011} on the European Southern Observatory (ESO) Very Large Telescope (VLT) in the context of the XQ-100 Legacy Survey \citep{Lopez2016}.
P16 showed that N{\sevensize V} is a key line to study the effects of the quasar ionization field, offering an excellent statistical tool for identifying outflow/intrinsic candidate NALs. Indeed, most of the N{\sevensize V} systems in our sample exhibit distinctive signatures of their intrinsic nature with respect to intervening NALs (described above), and N{\sevensize V}/C{\sevensize IV} column density ratios larger than 1 (see \citealp{FR09}).

The large number of Ly$\alpha$ lines characterizing the forest in the spectra of quasars at z$\rm_{em}$ = 3.5 - 4.72 prevents us in P16 from searching for individual N{\sevensize V} lines at large velocity offsets (we could reliably identify N{\sevensize V} only within 5000 km s$^{-1}$ of z$\rm_{em}$). Most of the Ly$\alpha$ forest is associated with moderate overdensities and traces filamentary structure on large scales, but some strong forest absorbers along with Lyman-limit systems (LLSs, N(H{\sevensize I}) $\geq$ 10$^{17.2}$ cm$^{-2}$) and damped Lyman alpha absorptions (DLAs, N(H{\sevensize I}) $\geq$ 10$^{20.3}$ cm$^{-2}$), are thought to be associated with galaxies and the circumgalactic medium (CGM; e.g. \citealp{Faucher2011, Fumagalli2011}). Therefore, intervening metals associated with Ly$\alpha$ absorbers can probe different environments: from the ISM to the outer regions of galaxies (the CGM) far from the quasar to the more diffuse intergalactic medium (IGM).

In this work, we apply the stacking technique to the XQ-100 spectra to look for N{\sevensize V} at large velocity separations from z$\rm_{em}$. Indeed, when an ensemble of independent sight lines is stacked to produce a composite spectrum, the numerous stochastic H{\sevensize I} absorptions average together and the resulting spectrum is flat, revealing any strong metal signal buried within the Ly$\alpha$ forest. These measurements will complement our previous study on N{\sevensize V} and test the robustness of our claim on the use of N{\sevensize V} as a criterion to select intrinsic NALs.

The work is organized as follows: Section~\ref{civ_sample} describes our methodology for identifying NALs; Section~\ref{stacking} collects the details of the procedure followed to build the stacking spectra. Our results are presented in Section~\ref{results} and Section~\ref{discussion} discusses the original findings of our study. Our conclusions are summarized in Section~\ref{conclusions}.  

We adopt a $\Lambda$CDM cosmology throughout this manuscript, with $\rm \Omega_M$ = 0.315, $\Omega_{\Lambda}$ = 0.685, and H$_0$ = 67.3 km s$^{-1}$ Mpc$^{-1}$ \citep{planck14}.

\section{DATA}
 \label{civ_sample}

To carry out the study of N{\sevensize V}  absorption systems, we start from the C{\sevensize IV}  sample that we collected in P16. The C{\sevensize IV}  NALs are identified in quasar spectra from the XQ-100 Legacy Survey \citep{Lopez2016}.
The most common metal transition found in quasar spectra is C{\sevensize IV}. Roughly half of the Ly$\alpha$ forest with N(H{\sevensize I}) $\geq$ 10$^{14.5}$ cm$^{-2}$ is C{\sevensize IV}-enriched at z $\sim$ 3 \citep{Cowie1995, Songaila1998}. C{\sevensize IV} enrichment has even been suggested at lower N(H{\sevensize I}) \citep{Songaila1998, Ellison2000, Schaye2003, Dodorico2016}. This is the main reason why we start our analysis from a sample of C{\sevensize IV} absorbers, as well as a better alignment of the C{\sevensize IV} systems with other metal lines with respect to Ly$\alpha$ absorbers \citep{Kim2016}.

In P16, we identify almost one thousand C{\sevensize IV} doublets (1548.204, 1550.78 \AA \footnote{ Wavelengths and oscillator strengths used in this work are taken from \citet{Morton2003}.}) covering the redshift range 2.55 < z < 4.73. The catalogue is produced with the goal of mapping the incidence of NALs in the quasar rest-frame velocity space. Our study includes any C{\sevensize IV} absorber outside the Ly$\alpha$ forest in each spectrum (up to $\sim$ 71,580 km s$^{-1}$ from z$\rm_{em}$). We refer to systems occurring at less than 5000 km s$^{-1}$ from the systemic redshift of the quasar as {\it associated} absorption lines (AALs).

We refer to P16 for a complete description of the identification and measurement of equivalent width (EW) and column density (N) of C{\sevensize IV} absorbers. 

The absorber velocity (v$\rm_{abs}$) with respect to the quasar systemic redshifts is computed by the relativistic Doppler formula,

\begin{equation}
\\\\\\\ \rm \beta \equiv \frac{v_{abs}}{c} =  \frac{(1+z_{em})^2 - (1+z_{abs})^2}{(1+z_{em})^2 + (1+z_{abs})^2}
\label{v_abs}
\end{equation}

\begin{raggedleft}{where z$\rm_{em}$ and z$\rm_{abs}$ are the emission redshift of the quasar and the absorption redshift of the NAL, respectively, and c is the speed of light.}\end{raggedleft}

Our final sample includes 986 C{\sevensize IV} doublets with $-$1000 < v$\rm_{abs}$ < 71,580 km s$^{-1}$ and equivalent widths 0.015 \AA\,< EW < 2.00\,\AA. 

Thanks to the high S/N and wavelength extent of our spectra, we are able to search for other common ions (N{\sevensize V}, Si{\sevensize IV}, C{\sevensize II}, etc) at the same redshifts of the detected C{\sevensize IV} absorbers. However, contamination by the Ly$\alpha$ forest prevents us from investigating the same velocity range for all the ions. In particular, our final sample consists of 46 N{\sevensize V} absorbers with column density down to Log N(N{\sevensize V}) $\sim$ 12.8 cm$^{-2}$, corresponding to a probability of 38 per cent to find a N{\sevensize V} at the same z$\rm_{abs}$ of a given C{\sevensize IV} NAL with v$\rm_{abs}$ < 5000 km s$^{-1}$. 55 per cent of N{\sevensize V} systems have column densities larger than Log N(N{\sevensize V}) > 14 cm$^{-2}$ and 85 per cent have values larger than Log N(N{\sevensize V}) > 13.5 cm$^{-2}$. Furthermore, 68 per cent of the N{\sevensize V} NALs have column densities larger than the corresponding C{\sevensize IV}.

\section{STACKING PROCEDURE}
\label{stacking}

We adopt the stacking technique to explore the presence of N{\sevensize V} signal in the Ly$\alpha$ forest associated with the C{\sevensize IV} absorbers described in the previous section.
The first step to produce the stacked spectrum is the continuum normalization of all the XQ-100 quasar spectra. 
The continuum placement in the Ly$\alpha$ forest is highly subjective, due to few portions of the spectrum free from absorption (e.g. \citealp{Kirkman2005}). The continuum is particularly difficult to identify at the position of DLAs. 
The estimate of the continuum level is done manually by selecting points along the quasar continuum free of absorption (by eye) as knots for a cubic spline. For all sightlines, the continuum placement is inspected visually and adjusted such that the final fit resides within the variations of regions with clean continuum. Then, the flux and error arrays are continuum-normalized.
The accuracy of the fits is as good as or better than the S/N of these clean continuum regions (see \citealp{Lopez2016} for a detailed description of the adopted procedure). 

We follow the subsequent scheme for the construction of the average spectrum.

(i) For each of the 986 C{\sevensize IV} systems identified in P16, we consider the position of the velocity component with the highest optical depth of the absorption profile as the redshift of the system, z$\rm_{abs}$, around which we select the spectrum region to stack. For completeness, we will show how the results change calculating the barycentre position of the system weighting the wavelengths with the optical depth of the line profile (see Sec.~\ref{barycentre}). 

(ii) We use the C{\sevensize IV} centre, z$\rm_{abs}$, to select the region of the spectrum where the corresponding N{\sevensize V} doublet should be located. Specifically, we select a region of $\pm$ 1000 km s$^{-1}$ around the expected position of the N{\sevensize V} centre (for both doublet components; i.e. N{\sevensize V}$\lambda$1238 \, and N{\sevensize V}$\lambda$1242). 

(iii) Each selected region is shifted to the corresponding z$\rm_{abs}$ rest-frame.

(iv) We generate a rest-frame wavelength array with fixed wavelength step $\Delta\lambda$. The step value is set to be slightly larger than the pixel size of the UV spectra. The visible arm of X-shooter has a mean resolution of $\lambda$/$\Delta\lambda$ = 8800, while in the UV it goes down to $\sim$ 5000. Thus, a region with the same extent is sampled with a larger number of pixels in the visible than in the UV region of the spectrum.

(v) For each selected spectral region, we compute the contribution of each pixel to the cells of the final grid. 

(vi) All the flux values in each cell of the final grid are then averaged to produce the stacked spectrum.

(vii) Uncertainties on the observed stack are estimated through the bootstrap resampling technique. From a given number, {\it N}, of rest-frame spectral regions, we consider the first pixel of each region as the element of an array. We randomly sample from this array, allowing for replacement, to form a new sample (called bootstrap sample) that is also of size {\it N}. This process is repeated a large number of times (e.g. 1000 times), and for each of these bootstrap samples we compute its mean. We now have a histogram of bootstrap means. This provides an estimate of the shape of the distribution of the mean. The measured standard deviation of this distribution is the error we associate with the first pixel of the stacked spectrum. Then, the method is applied to all the pixels of the rest-frame spectral regions to be stacked.    

Our study is not so sensitive to the continuum placement because we do the analysis relative to the average IGM opacity, without the goal of measuring the opacity.

For each sightline, we mask out the regions that have clear strong absorption features (i.e. Lyman breaks, sub-DLAs with N(H{\sevensize I}) $\sim$ 10$^{19-20.3}$ cm$^{-2}$ and DLAs) when they coincide with those to be stacked. In particular, we take advantage of the DLA identification done by \citet{Ramirez2016}. We define the extremes of the DLA absorption at the wavelengths where the wings of the DLA again match the level of the continuum to within the values of the flux error array.

We exclude 77 N{\sevensize V} regions because they completely or partially overlap with Lyman breaks or DLA and sub-DLA absorptions. Our final sample consists of 909 N{\sevensize V} selected regions, the centre of which is within $-$1000 < v$\rm_{abs}$ < 71,580 km s$^{-1}$ of z$\rm_{em}$. We extend the explored velocity range to $-$1000 km s$^{-1}$  to take uncertainties in the quasar systemic redshift into account \citep{Lopez2016} and to consider possible inflows.

This method is powerful in detecting even weak absorption features in the Ly{\sevensize $\alpha$} forest, where confusion noise from forest absorption lines makes it very difficult to identify individual metal lines reliably. However, N{\sevensize V} is one of the most rarely detected ions in optically thin (i.e. N(H{\sevensize I}) $\lesssim$ 10$^{17}$ cm$^{-2}$) intergalactic absorption systems and its signal is diluted if we stack regions selected within any velocity shift from z$\rm_{em}$ up to 71,580 km s$^{-1}$.

\section{RESULTS}
\label{results}

The following sections collect the results of the search for metal absorptions corresponding to C{\sevensize IV} clouds with 10$^{11.5}$ < N(C{\sevensize IV}) < 10$^{16.5}$ cm$^{-2}$ (see P16). We first investigate the presence of a N{\sevensize V} signal as a function of the velocity shift from the quasar to test the nature of N{\sevensize V}. Then, we examine many other lower and higher ionization species. The main goal is to characterize the ionization state of the gas both close and far from the continuum source.

\subsection[]{N{\sevensize \bf V} Absorption}
\label{NV}

\begin{figure*}
 \includegraphics[width=\columnwidth]{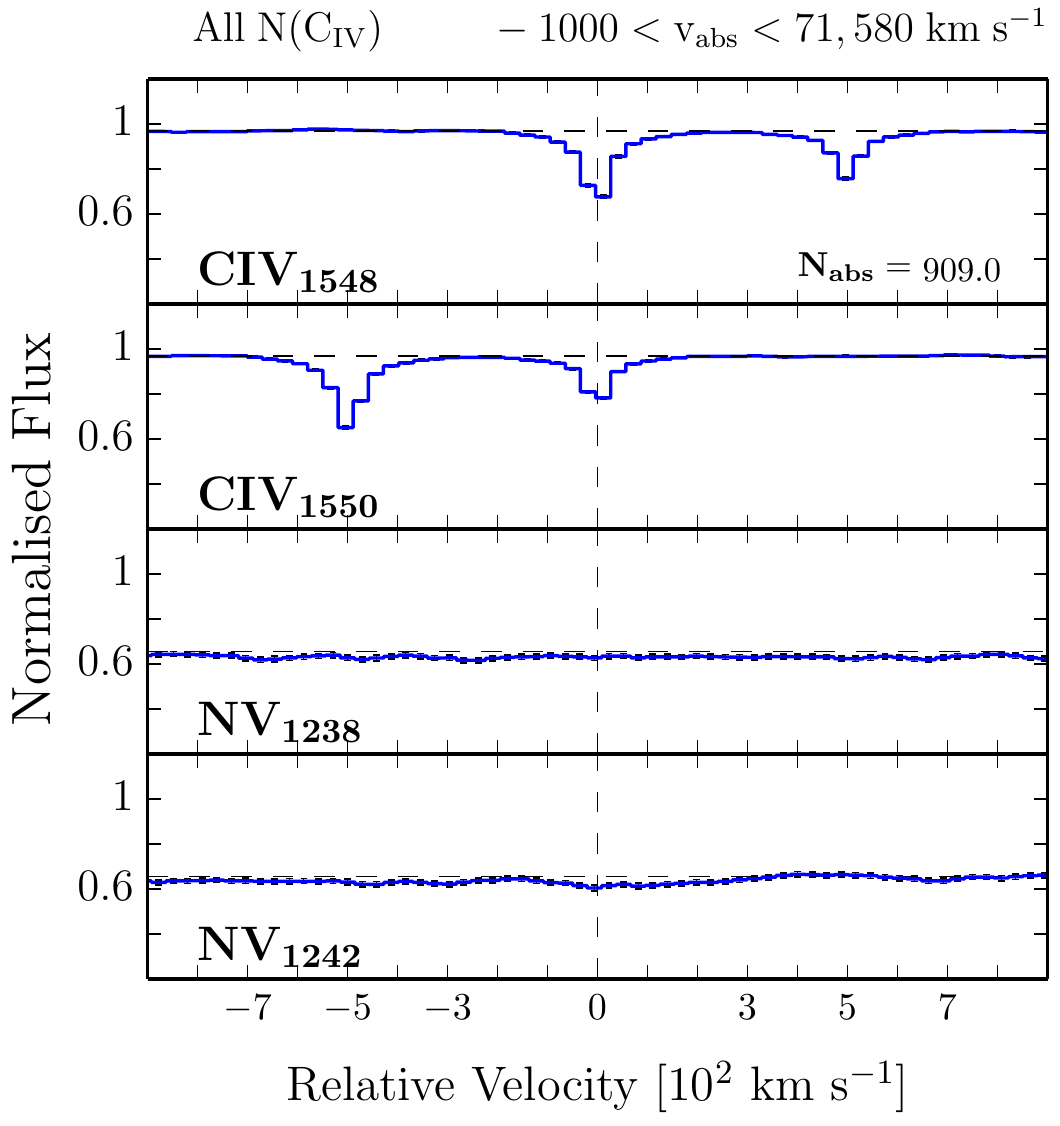}
 \includegraphics[width=\columnwidth]{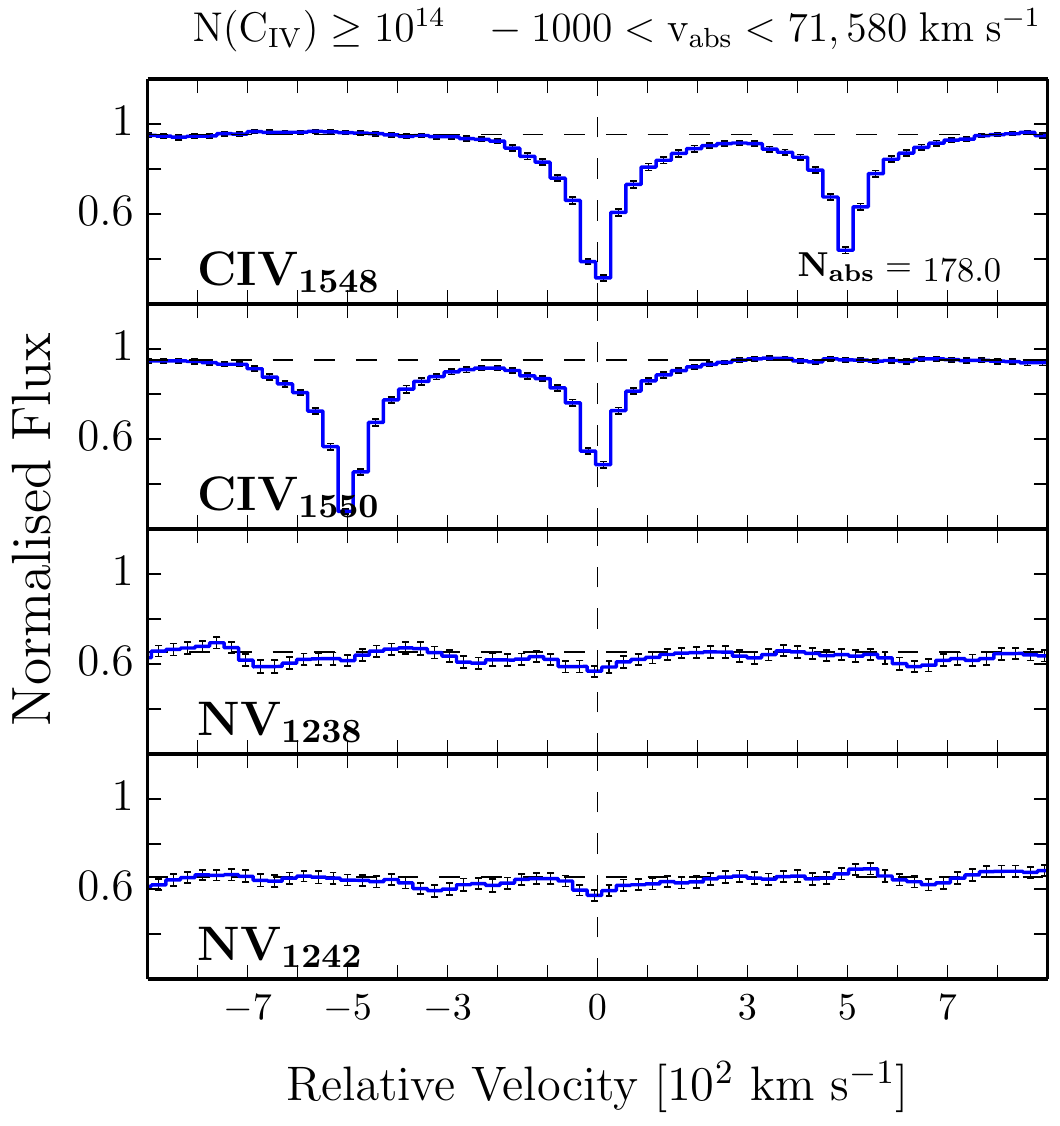}
 \caption{Left: average absorption line spectra for C{\sevensize IV} and N{\sevensize V}, selected within any velocity separation from z$\rm_{em}$. $\rm N_{abs}$ indicates the number of absorbers considered to build the stacked spectrum. Right: same as for the left panels, but in this case only regions corresponding to the strongest C{\sevensize IV} NALs (i.e. N(C{\sevensize IV}) $\geq$ 10$^{14}$ cm$^{-2}$) are included in the stack. Error bars are 1$\sigma$ and are estimated by bootstrapping the data using 1000 realizations.}
 \label{fig:civnv_0-75000}
\end{figure*}

\begin{figure*}
 \includegraphics[width=\columnwidth]{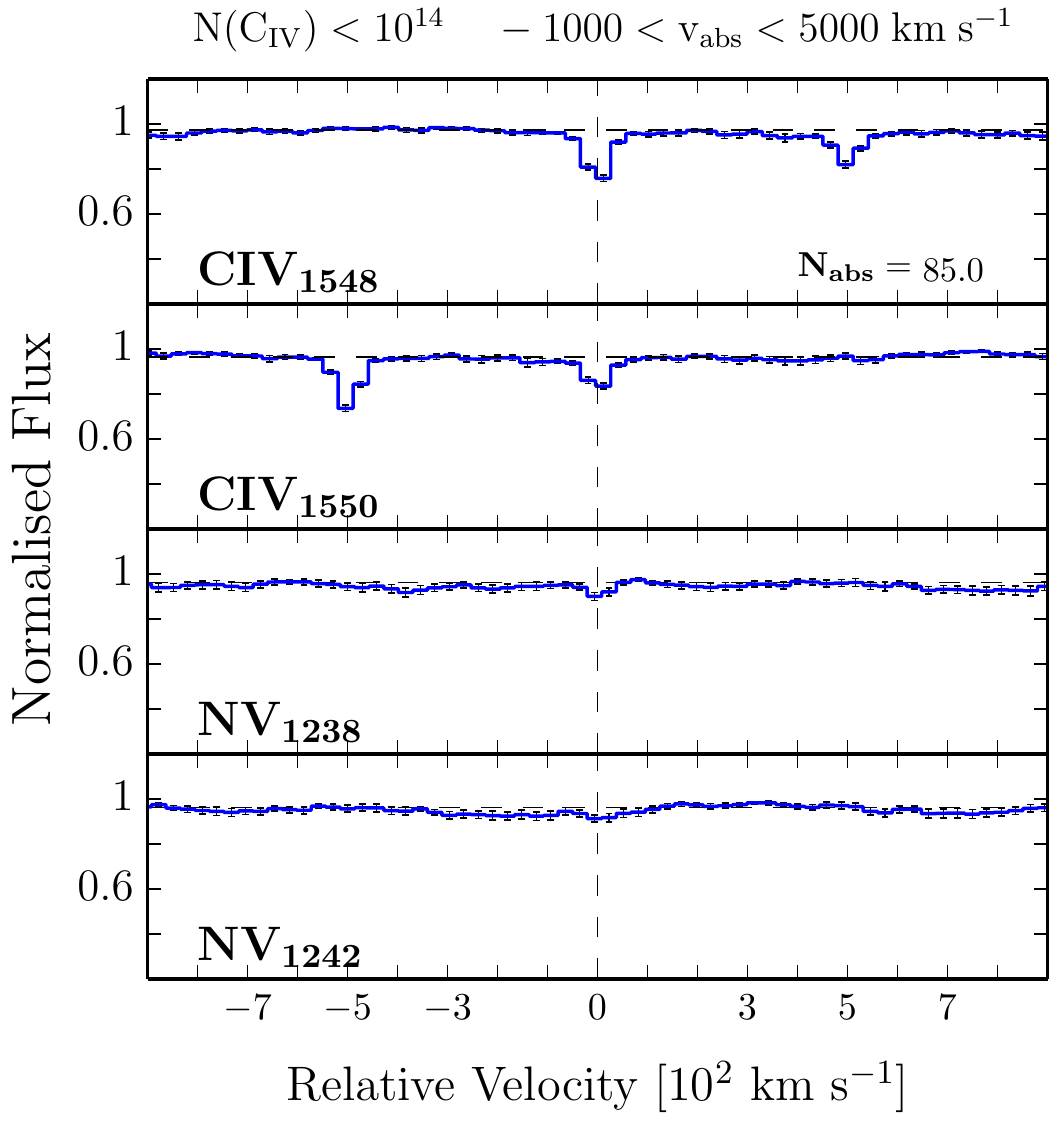}
  \includegraphics[width=\columnwidth]{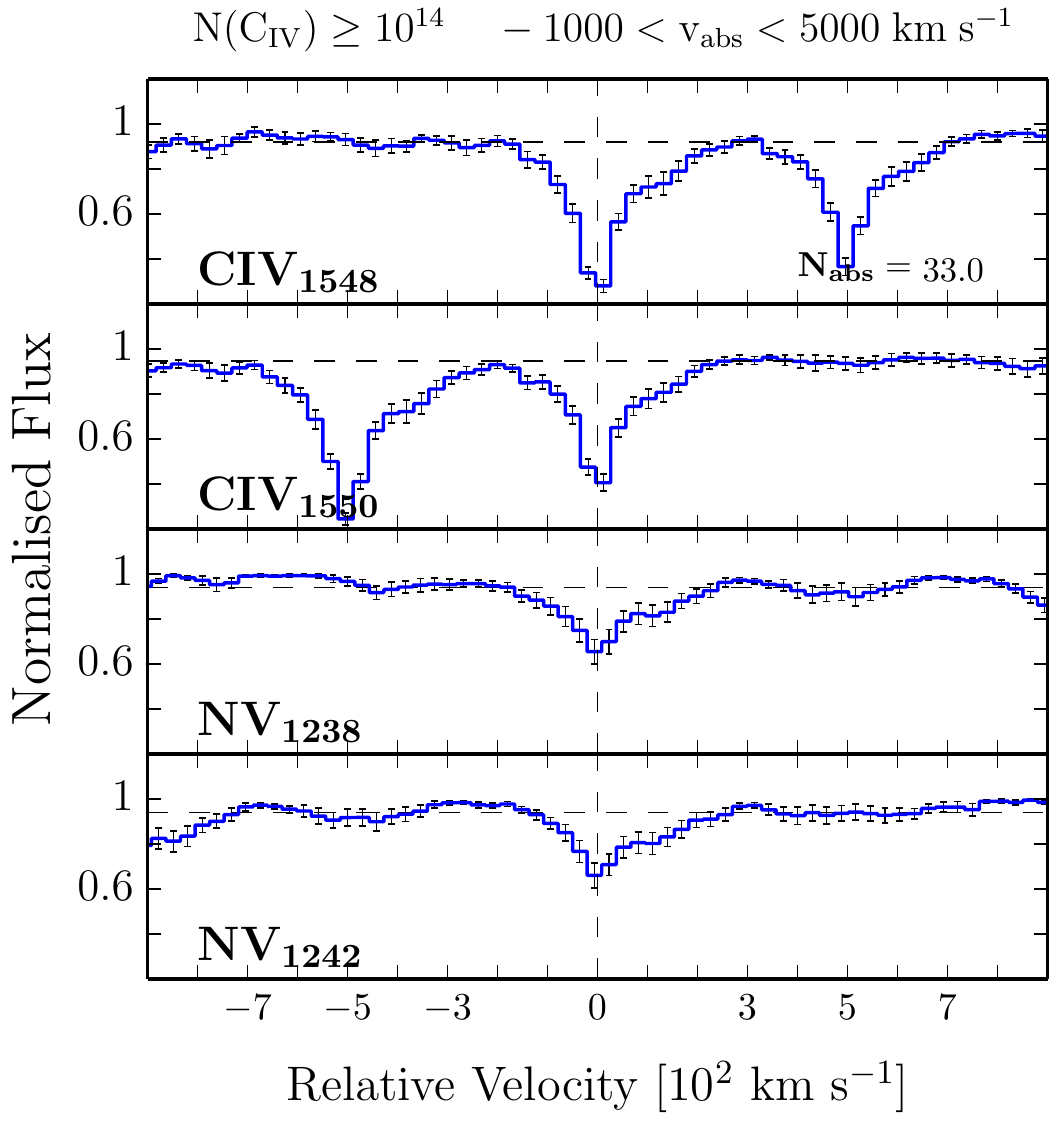}

 \caption{Left: average absorption line spectra for C{\sevensize IV} and N{\sevensize V} selected within 5000 km s$^{-1}$ of z$\rm_{em}$. Only regions corresponding to C{\sevensize IV} NALs with N(C{\sevensize IV}) < 10$^{14}$ cm$^{-2}$ are included in the stack. $\rm N_{abs}$ indicates the number of absorbers considered to build the stacked spectrum. Right: same as for the left panels, but in this case N(C{\sevensize IV})$\geq$ 10$^{14}$ cm$^{-2}$ is considered as threshold for the stacking. Error bars are 1$\sigma$ and are estimated by bootstrapping the data using 1000 realizations.}
 \label{fig:civnv_0-5000}
\end{figure*}

The left panel of Fig.~\ref{fig:civnv_0-75000} shows the results for C{\sevensize IV} and N{\sevensize V} absorbers obtained by stacking the whole sample of sight lines. The zero velocity point, v = 0 km s$^{-1}$, represents the z$\rm_{abs}$ of the deepest component of the corresponding C{\sevensize IV} system. The average spectrum appears quite flat for both components of the N{\sevensize V} doublet, as expected in absence of a strong signal, due to the stochastic nature of the Ly$\alpha$ forest. The continuum level is decreased to a value below 1 due to the cumulative effect of the Ly$\alpha$ forest absorption.

The right panel of Fig.~\ref{fig:civnv_0-75000} exhibits the C{\sevensize IV} and N{\sevensize V} composite spectra derived from the stacking of the regions corresponding to C{\sevensize IV} NALs with N(C{\sevensize IV}) $\geq$ 10$^{14}$ cm$^{-2}$. This column density threshold matches the value we use in P16 to distinguish between weak and strong C{\sevensize IV} systems. Selecting the strongest systems in the sample, we see only a small hint of coherent absorption features in the profile of the two N{\sevensize V} doublet components. Fig.~\ref{fig:civnv_0-75000} combines all of the NALs regardless of velocity shift. In the following, we investigate the presence of a N{\sevensize V} signal as a function of the velocity separation from the quasar to understand when it becomes significant.
To this end, we produce N{\sevensize V} composite spectra in bins of velocity offset from z$\rm_{em}$.

Fig.~\ref{fig:civnv_0-5000} shows the stacked spectra of C{\sevensize IV} and N{\sevensize V} absorbers with $-$1000 < v$\rm_{abs}$ < 5000 km s$^{-1}$ and as a function of the corresponding N(C{\sevensize IV}). In particular, in the left panels the stacked regions are selected so that N(C{\sevensize IV}) < 10$^{14}$ cm$^{-2}$, while in the right ones N(C{\sevensize IV}) $\geq$ 10$^{14}$ cm$^{-2}$. 
The first column density threshold chosen for the stacking produces a hint of coherent absorption in the profile of both the N{\sevensize V} components (left panels). The signal of the absorption line is weak, but represents a 4 $\sigma$ detection. The N{\sevensize V} composite spectrum shows the N{\sevensize V} lines clearly present at > 15$\sigma$ confidence once only the regions corresponding to the strongest C{\sevensize IV} are selected to be stacked (right panels). We note that both N{\sevensize V} lines display the same kinematic profile and they also match those of the corresponding C{\sevensize IV} NAL.

\begin{table}
\caption[]{Probability that an individual C{\sevensize IV} NAL within 5000 km s$^{-1}$ of z$\rm_{em}$, shows a N{\sevensize V} absorber at the same absorption redshift, as a function of its column density.}
\begin{center}
\label{tab_num_CIV}
\begin{tabular}{ c  c c  }
\hline
\hline

   & Number of C{\sevensize IV} $\rm^{a}$  &    C{\sevensize IV}  with N{\sevensize V} $\rm^{b}$ \\

 \hline
all& 122& 38\%\\

Log N(C{\sevensize IV}) $\geq$ 14.3& 25& 72\%\\

Log N(C{\sevensize IV})  $\geq$ 14& 42& 55\%\\
Log N(C{\sevensize IV})  $\geq$ 13.5& 85& 44\%\\

Log N(C{\sevensize IV})  < 14.3& 97& 29\%\\

Log N(C{\sevensize IV})  < 14& 80& 29\%\\

\hline
\\
\end{tabular}
\end{center}
$\rm^{a}$ Number of C{\sevensize IV} NALs with v$\rm_{abs}$< 5000 km s$^{-1}$ and with column density in the selected range\\
$\rm^{b}$ Fraction of C{\sevensize IV} NALs with a N{\sevensize V} absorber at the same absorption redshift. Detection limit on N{\sevensize V} is  Log N(N{\sevensize V}) $\sim$ 12.8 cm$^{-2}$
\end{table}

\begin{figure*}
 \includegraphics[width=\textwidth]{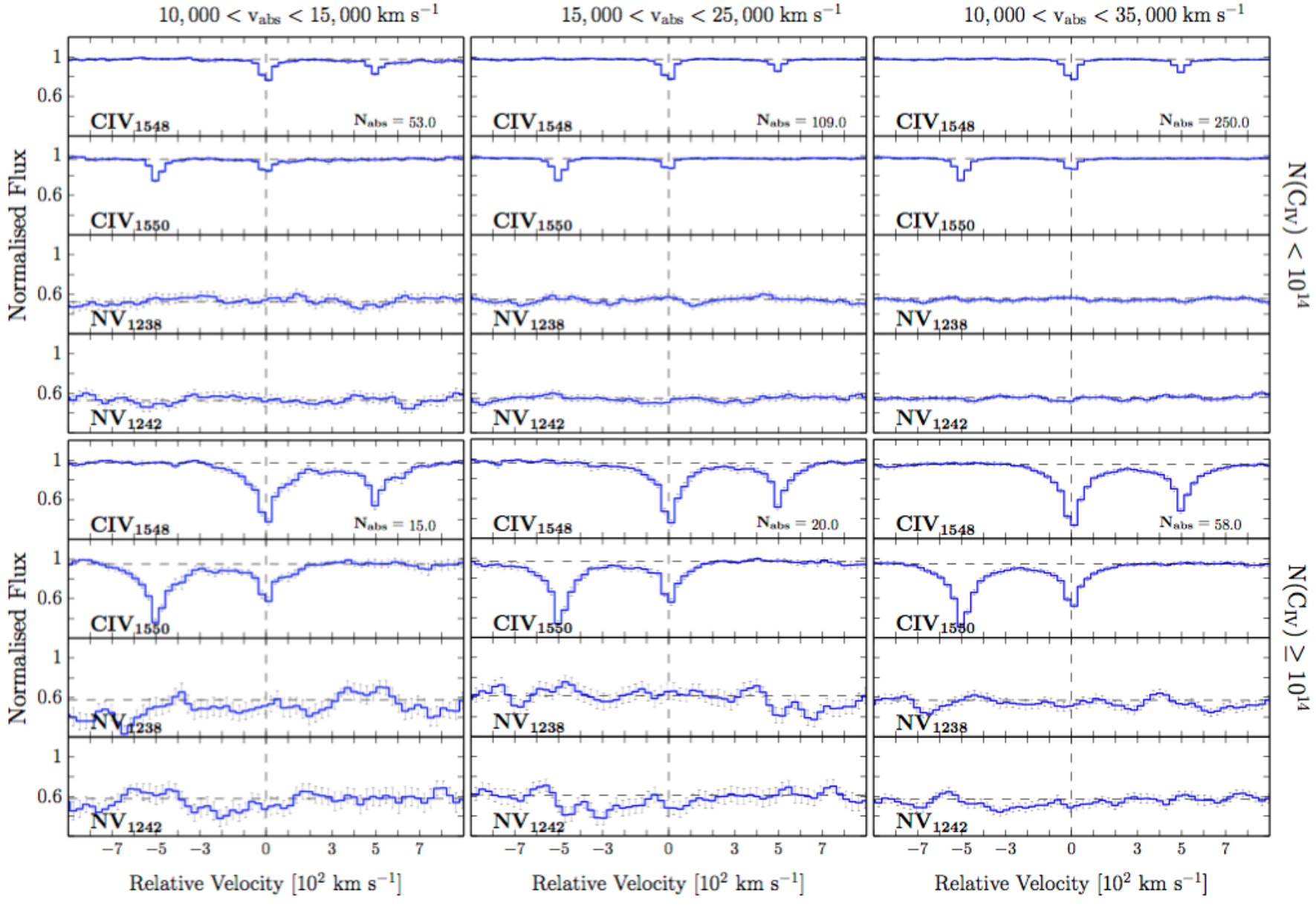}

 \caption{Top panels: the average absorption line spectra produced selecting only C{\sevensize IV} systems with N(C{\sevensize IV})$\leq$ 10$^{14}$ cm$^{-2}$ are shown in different bin size of velocity offset from the z$\rm_{em}$. Bottom panels: same as for the top panels, but with a different column density threshold, N(C{\sevensize IV})$\geq$ 10$^{14}$ cm$^{-2}$.}
 \label{fig:multi}
\end{figure*}

The associated region is the only spectral window in which we are able to identify individual N{\sevensize V} NALs. 
The number of C{\sevensize IV} doublets within 5000 km s$^{-1}$ of z$\rm_{em}$ as a function of the column density is reported in Tab.~\ref{tab_num_CIV}, as well as, the fraction of C{\sevensize IV} showing associated N{\sevensize V}. 
The stronger the C{\sevensize IV} column density, the higher the probability of selecting systems with N{\sevensize V} at the same redshift. This trend is not due to a detection bias affecting the identification of N{\sevensize V} NALs at low column densities. Indeed, the high and uniform S/N of the XQ-100 spectra allows us to identify N{\sevensize V} systems reliably down to Log N(N{\sevensize V}) $\sim$ 12.8 cm$^{-2}$ (see P16). In addition, associated N{\sevensize V} NALs are preferentially stronger than the corresponding C{\sevensize IV}  \citep{FR09}.
However, we adopt N(C{\sevensize IV}) = 10$^{14}$ cm$^{-2}$ as the column density threshold to distinguish between weak and strong systems. This restriction is motivated by the requirement to have a minimum number of systems within a given velocity separation from z$\rm_{em}$ to build the composite spectrum. This is crucial, especially when we investigate large velocity offsets from z$\rm_{em}$, where the number density of C{\sevensize IV} NALs exhibits a clear drop with respect to the excess of lines found in proximity to the quasar (see P16).
Moreover, \citet{FR09} have shown that intervening N{\sevensize V} systems are systematically weaker than the associated ones: no intervening N{\sevensize V} system with N(N{\sevensize V}) $\geq$ 10$^{14}$ cm$^{-2}$ has been detected. It is therefore very important to explore the regime of weak lines.

To investigate the N{\sevensize V} signal beyond the associated region, we build the N{\sevensize V} composite spectra in bins of 5000 km s$^{-1}$ up to 71,580 km s$^{-1}$ from z$\rm_{em}$ both for weak and strong C{\sevensize IV} systems.
In particular, Fig.~\ref{fig:multi} (left column) shows the results for C{\sevensize IV} and N{\sevensize V} absorbers with 10,000 < v$\rm_{abs}$ < 15,000 km s$^{-1}$. The stacked spectra produced selecting only C{\sevensize IV} systems with N(C{\sevensize IV}) $\leq$ 10$^{14}$ cm$^{-2}$ are presented in the top panel, while in the bottom one only C{\sevensize IV} systems with N(C{\sevensize IV}) $\geq$ 10$^{14}$ cm$^{-2}$ are considered. In both case, there is no imprint of any significant coherent absorption feature in the profile of the two N{\sevensize V} doublet components. We show just one of the many 5000 km s$^{-1}$ velocity bins studied, because they all exhibit the same behavior. 
Fig.~\ref{fig:15-75_nodla} in the Appendix demonstrates that the stacking technique applied to the XQ-100 spectra is actually able to detect absorption lines in the Ly$\alpha$ forest. 

We then consider larger velocity bins, in order to increase any possible signal of intervening N{\sevensize V} absorbers. This is particularly important for the strong systems, which are much rarer at large separations from the quasar (see P16). The stacked spectra produced selecting only C{\sevensize IV} systems with 15,000 < v$\rm_{abs}$ < 25,000 km s$^{-1}$ (10,000 < v$\rm_{abs}$ < 35,000 km s$^{-1}$) are presented in the middle column of Fig.~\ref{fig:multi} (right column of Fig.~\ref{fig:multi}).

The absence of statistical significant intervening N{\sevensize V} absorption is a common feature independently of the parameters (N(C{\sevensize IV} ) and v$\rm_{abs}$) selected as threshold of the stacking. The mean rest frame EW upper limit found for the N{\sevensize V} absorption line in the different velocity bins explored is $\sim$ 0.03 \AA. 
This finding strongly suggests that a proximity of 5000 km s$^{-1}$ from the quasar is required for N{\sevensize V} to be detected.

\subsection{Other metal species}
\label{other_metals}

\begin{figure*}
 \includegraphics[width=\textwidth]{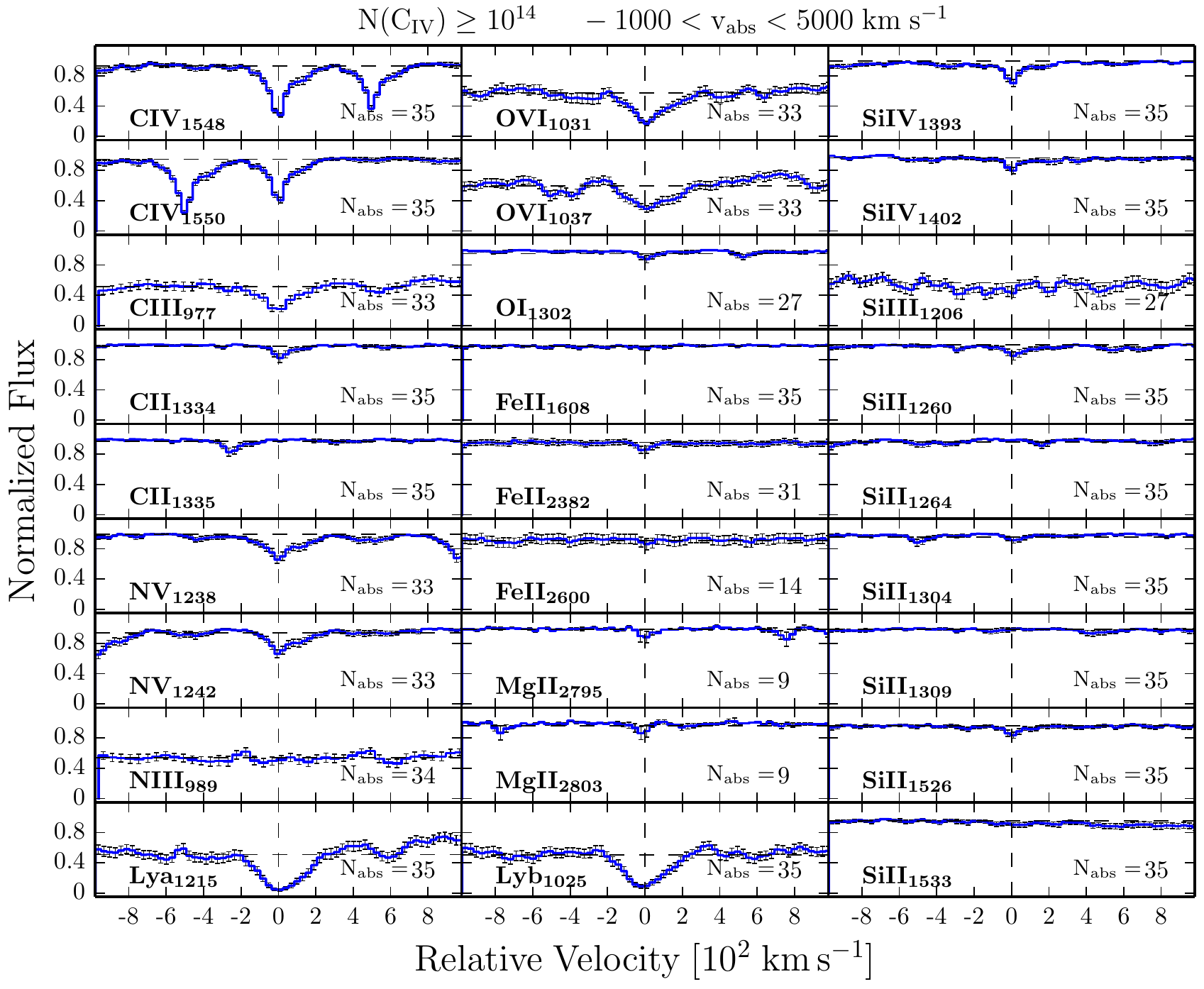}

 \caption{Composite spectra for the ions studied in this work, based on the C{\sevensize IV} systems with N(C{\sevensize IV}) $\geq$ 10$^{14}$ cm$^{-2}$ and $-$1000 < v$\rm_{abs}$ < 5000 km s$^{-1}$. Every panel reports the number of absorbers considered to build the stacked spectrum. N$\rm_{abs}$ is not the same for all the ion species, because each one undergoes a different masking for DLA, sub-DLA and Lyman breaks. The error bars are 1$\sigma$ and are estimated by bootstrapping the data using 1000 realizations. We include Mg{\sevensize II} and Fe{\sevensize II} $\lambda$ 2600 lines despite they both suffer of poor statistics.}
 \label{fig:0-5}
\end{figure*}

\begin{figure*}
 \includegraphics[width=\columnwidth]{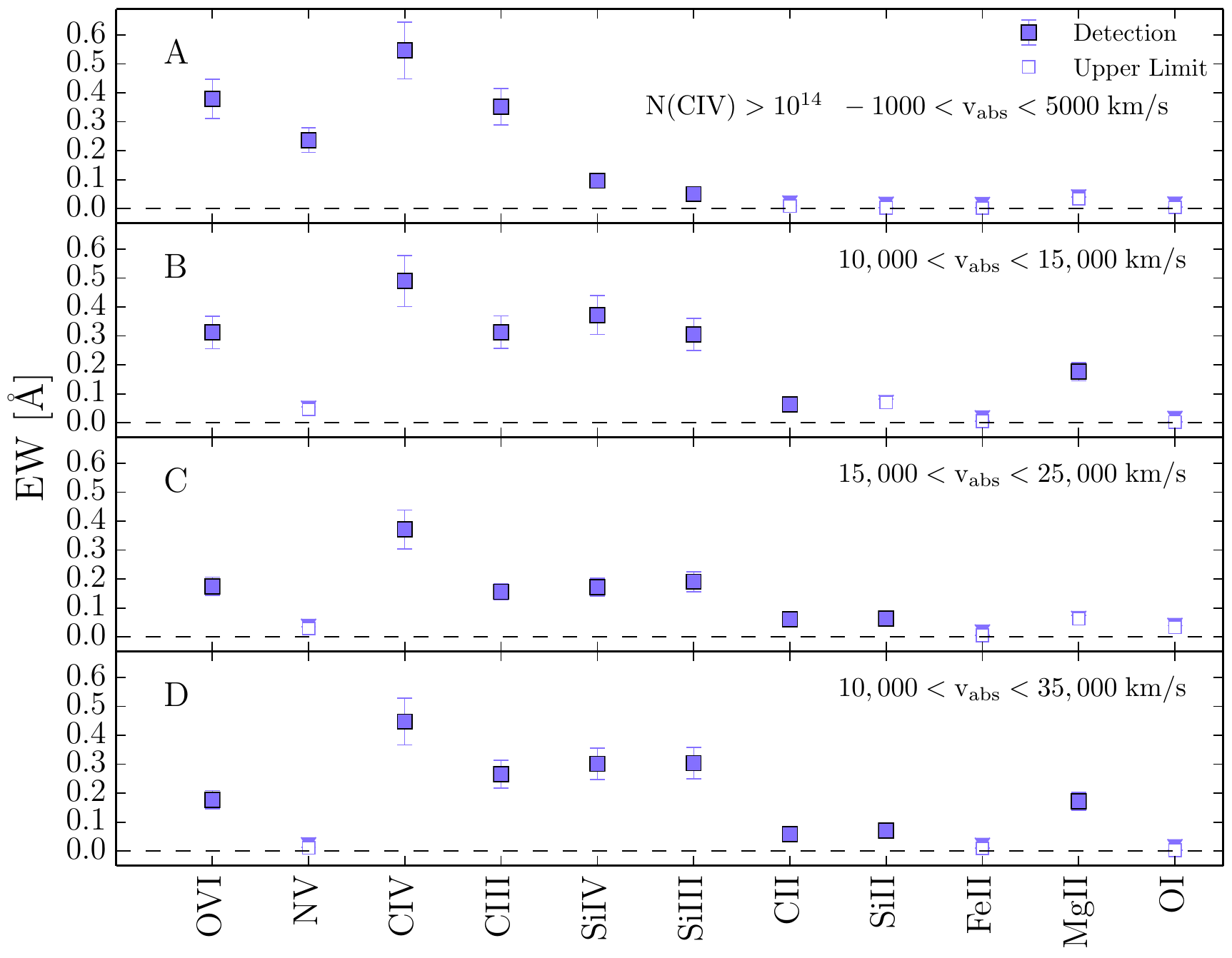}
 \includegraphics[width=\columnwidth]{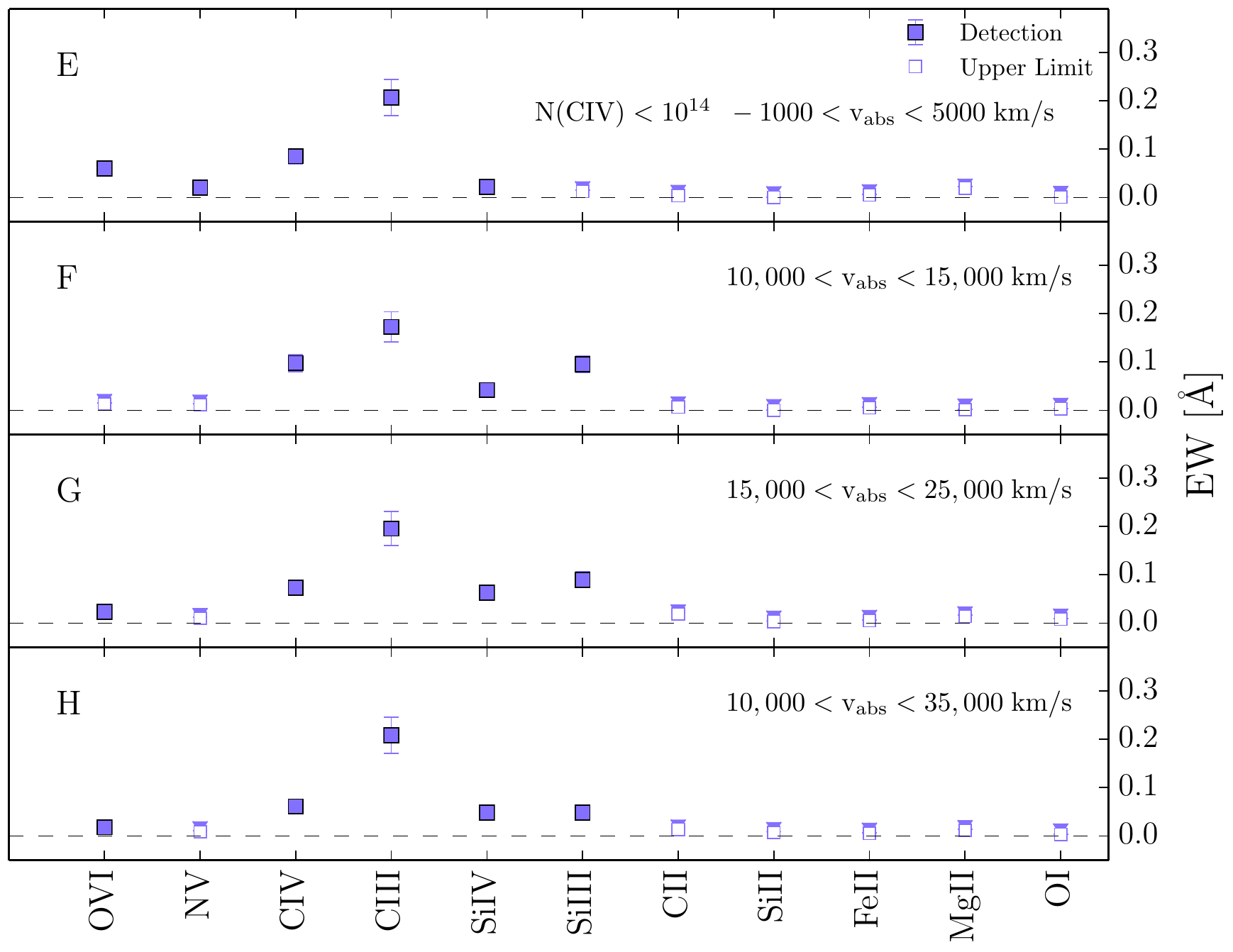}

 \caption{ Left: equivalent widths of metal species measured in order of decreasing ionization potential for composite spectra based on the C{\sevensize IV} systems with N(C{\sevensize IV})$\geq$ 10$^{14}$ cm$^{-2}$. Each panel displays an increasing velocity offset from z$\rm_{em}$ and bin size, going from top to bottom. C{\sevensize IV} DLAs have been excluded form the stacking. Right: same as for the left panels, but in this case C{\sevensize IV} systems with N(C{\sevensize IV})< 10$^{14}$ cm$^{-2}$ are included in the stack. The EW refers only to the strongest member of the doublets. Error bars are 1$\sigma$ and are estimated by bootstrapping the data using 1000 realizations.}
 \label{fig:EW_ions}
\end{figure*}

Besides the stacked spectra to detect N{\sevensize V} absorptions, we have computed with the same technique the stacked spectra for the following transitions associated with the sample of detected C{\sevensize IV} lines: Ly$\alpha$ $ \lambda$1215.67 \AA, Ly-$\beta$ $ \lambda$1025.72 \AA, C{\sevensize III} $ \lambda$977.02 \AA, C{\sevensize II} $ \lambda$1334.53 \AA, O{\sevensize I} $ \lambda$1302.16 \AA, Si{\sevensize III} $ \lambda$1206.50 \AA, Si{\sevensize II} $ \lambda$1260.42, 1304.37, 1526.70 \AA, Fe{\sevensize II} $ \lambda$ 2382.76 \AA, absorption and Si{\sevensize IV} $ \lambda \lambda$1393.76, 1402.77 \AA, O{\sevensize VI} $ \lambda \lambda$1031.92, 1037.61 \AA\,and Mg{\sevensize II} $ \lambda \lambda$2796.35, 2803 \AA\,doublet absorption. 

The composite spectrum based on the C{\sevensize IV} systems with N(C{\sevensize IV}) $\geq$ 10$^{14}$ cm$^{-2}$ and $-$1000 < v$\rm_{abs}$ < 5000 km s$^{-1}$ is shown in Fig.~\ref{fig:0-5}.  
Based on ionization arguments, we would expect that C{\sevensize IV} NALs showing associated ions like N{\sevensize V} or O{\sevensize VI} ought not to have low ionization transitions. To remove the effects of DLAs on our investigation, we exclude all C{\sevensize IV} associated with DLAs \citep{Ramirez2016} from the stacking. As a consequence, we do not see any significant evidence of absorption trough in the spectra of C{\sevensize II}, Mg{\sevensize II}, Si{\sevensize II}, Fe{\sevensize II} and O{\sevensize I} (see Fig.~\ref{fig:0-5_nodla}). This result is expected, since we have already measured a C{\sevensize II} covering fraction of 10 per cent for absorbers with v$\rm_{abs}$ < 10,000 km s$^{-1}$ along the line of sight (P16), in contrast with what is observed in the transverse direction \citep{Prochaska2014}. This finding seems to be confirmed by the absence of a strong Mg{\sevensize II} signal in the proximity of the quasar. We will come back to this point  in Section~\ref{mg2}. 

To quantify the detections clearly present in the composite spectrum, we measure the equivalent widths of the absorption lines. The continuum is first established by fitting a cubic spline function to the spectral regions free of apparent absorption lines. Then, the rest-frame equivalent width is measured as described in Section~\ref{civ_sample}. We present our metal-line equivalent width measurements in Fig.~\ref{fig:EW_ions}. For those lines that are not detected at over 3 $\sigma$ significance level, we give the 3$\sigma$ upper limits on their equivalent widths. 

The left panels of Fig.~\ref{fig:EW_ions} show the equivalent widths derived for composite spectra based on C{\sevensize IV} systems with N(C{\sevensize IV}) $\geq$ 10$^{14}$ cm$^{-2}$. Each panel displays an increasing velocity offset from z$\rm_{em}$ going from top to bottom. In panels (c) and (d), we examine  larger velocity bins, to allow low-ionization ions to be sufficiently numerous to produce measurable absorptions. 
In the proximity of the quasar (Fig.~\ref{fig:EW_ions}a), we detect species with ionization potentials down to and including Si{\sevensize IV} and intermediate lines such as Si{\sevensize III}.
Moving farther from the associated window (Fig.~\ref{fig:EW_ions}b-d), we no longer see N{\sevensize V}; this result is insensitive to the bin size chosen, as already shown in section~\ref{NV}. However, we do detect O{\sevensize VI} independently of the velocity bin size and the velocity shift from z$\rm_{em}$. 

Looking at increasing velocity offsets from z$\rm_{em}$, we can see how the quasar radiation field affects the gas along the propagation axis of the outflows. Indeed, low-ionization species (like C{\sevensize II}, Si{\sevensize II}, Mg{\sevensize II}) are mainly present in our strongest C{\sevensize IV} absorption sample when we move farther from the associated region (Fig.~\ref{fig:EW_ions}b-d). In Fig.~\ref{fig:EW_ions}(a) we examine gas more directly influenced by the quasar radiation. Closer to the emission of the quasar, the radiation field is more intense and the gas is more highly ionized (assuming a similar gas density); it is thus reasonable to expect that the occurrence of the low-ionization lines decreases there.

The number of strong C{\sevensize IV} systems steeply decreases beyond 5000 km s$^{-1}$ from z$\rm_{em}$ \citep{Perrotta2016}. Thus, if we consider individual velocity bin size of 5000 km s$^{-1}$ along the line of sight, it can happen that some isolated line dominates the signal of a given ion (e.g. see the Si{\sevensize IV}, Si{\sevensize III} and Mg{\sevensize II} in Fig.~\ref{fig:EW_ions}b). In particular, Mg{\sevensize II} and Fe{\sevensize II} $\lambda$2600 are the ions that suffer the most poor statistics, due to their location in the near-IR. Indeed, their expected positions often coincide with deep atmospheric absorption bands that do not allow us to include those systems in the stacking. This is the main reason for exploring larger velocity bins. Increasing the statistics of systems that build the composite spectrum, we are more confident to catch the general behavior of the ions. When we consider larger bin size (Fig.~\ref{fig:EW_ions}c and d), we do detect low-ionization potential species such as C{\sevensize II}, Si{\sevensize II} and Mg{\sevensize II}, in agreement with previous works based on composite spectra (e.g. \citealp{Lu1993, Prochaska2010, Pieri2010, Pieri2014}). 

The right panels of Fig.~\ref{fig:EW_ions} show the equivalent widths derived for composite spectra based on the C{\sevensize IV} systems with N(C{\sevensize IV}) < 10$^{14}$ cm$^{-2}$. Each panel displays an increasing velocity offset from z$\rm_{em}$ and bin size, going from top to bottom. C{\sevensize IV} DLAs have been excluded from the stack. The C{\sevensize IV} sample exploited in this work shows a statistically significant excess within 10,000 km s$^{-1}$ of z$\rm_{em}$ with respect to the random incidence of NALs, which is especially visible for the weak lines \citep{Perrotta2016}. The observed excess includes contributions from the following: i) outflowing gas that is ejected from the central AGN, ii) intervening absorbers physically related to the galaxy halos that cluster with the quasar host and iii) environmental absorption, which arises in either the ISM of the quasar host galaxy or the IGM of the galaxy's group or cluster. Most likely, C{\sevensize IV} systems showing a N{\sevensize V} absorber at the same redshift are intrinsic. 
We do not detect low-ionization ions within the associated window, nor moving farther from z$\rm_{em}$ (Fig.~\ref{fig:EW_ions}f-h). This is true independently of the velocity bin size chosen. Low-ionisation metal transitions, such as Mg{\sevensize II} and C{\sevensize II}, are mostly found at N(H{\sevensize I}) $\geq$ 10$^{16}$ cm$^{-2}$ (e.g. \citealp{Farina2014}). Therefore, most of the weak C{\sevensize IV} systems in our sample are probably associated with N(H{\sevensize I})  < 10$^{16}$ cm$^{-2}$ (see also \citealp{Kim2016, Dodorico2016}). In addition, we detect O{\sevensize VI} only if we consider a velocity bin width of at least 10,000 km s$^{-1}$, although its signal is weak.

\subsection{MgII  covering fraction}
\label{mg2}

Various studies based on projected quasar pairs agree that quasar host galaxies at z > 1 exhibit a high incidence of optically thick, metal-enriched absorption systems traced by H{\sevensize I} Ly$\alpha$, C{\sevensize II}, C{\sevensize IV}  and Mg{\sevensize II}  absorption along the transverse direction at d $\lesssim$ 300 kpc, but low incidence along the foreground quasar sightline itself \citep{Bowen2006, Hennawi2006, Shen2012, Farina2013, Farina2014, Hennawi2013, Prochaska2014, Lau2016, Lau2017, Chen2018}. This contrast indicates that the ionizing emission from quasars is highly anisotropic, in qualitative agreement with the unified model of AGN \citep{Antonucci1993, Netzer2015}. 

In particular, \citealp{Farina2014} (and \citealp{Johnson2015}) characterized the cool gas contents across the line of sight of quasar host haloes as a function of projected distance and found a large amount
of Mg{\sevensize II} with EW > 0.3 \AA\, extending to 200 (300) kpc from the quasars considered. The steep drop in covering fraction, f$\rm_C^{Mg{\sevensize II} }$, at distances larger than 200 (300) kpc requires that this Mg{\sevensize II} gas lies predominantly within the host halo (see fig.4 of \citealp{Farina2014} or fig.3 of \citealp{Johnson2015}).

\begin{figure}
 \includegraphics[width=\columnwidth]{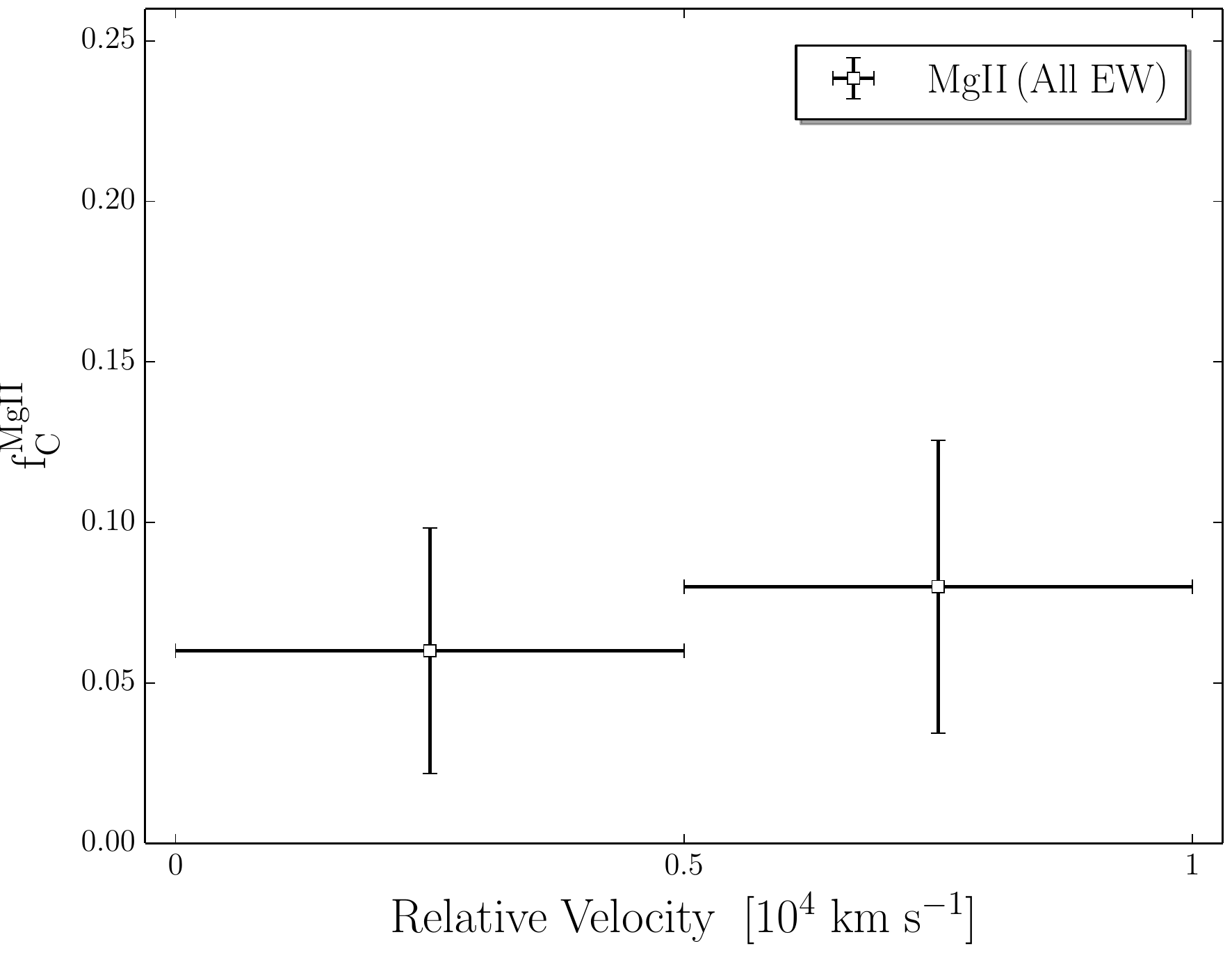}

 \caption{Covering fraction, f$\rm_C$, for Mg{\sevensize II} $\lambda$ 2796 estimated from the fraction of quasars exhibiting at least one absorber in bins of velocity offset from z$\rm_{em}$. The rest frame equivalent width detection limit (in the near-IR) in our sample is 0.036 \AA\, (see P16).}
 \label{fig:Mg2_covering}
\end{figure}

In contrast with these results, we have shown in Section~\ref{other_metals} the lack of a strong Mg{\sevensize II} absorption signal in the proximity of the quasar along the line of sight. We now measure the Mg{\sevensize II} covering fraction, f$\rm_C^{Mg{\sevensize II} }$, as a function of the velocity offset from z$\rm_{em}$. This is defined as the ratio between the number of quasars showing at least one Mg{\sevensize II} absorber within a given velocity separation of z$\rm_{em}$ and the total number of quasars with spectral coverage across the Mg{\sevensize II} lines. Poissonian uncertainties are taken into account. The result is shown in Fig.~\ref{fig:Mg2_covering}. The frequency with which Mg{\sevensize II} is detected in our survey is 6 per cent for absorbers with v$\rm_{abs}$ < 5000 km s$^{-1}$ and 8 per cent for those with 5000 < v$\rm_{abs}$ < 10,000 km s$^{-1}$. Moreover, if we consider only absorbers with v$\rm_{abs}$ < 5000 km s$^{-1}$ and EW > 0.3 \AA, we see that they are all related to identified DLA systems \citep{Berg2017}, probably unrelated to the quasar host galaxy. This supports our finding in P16 for a dearth of cool gas traced by C{\sevensize II} in AALs along direct lines of sight to the quasars, in contrast with results based on projected quasar pairs \citep{Prochaska2014, Lau2017}.

The detection fraction derived in \citealp{Farina2014} (or \citealp{Johnson2015}) cannot be compared directly with our result. Indeed, the velocity offset of each Mg{\sevensize II} absorber detected in our survey cannot be translated directly into a physical separation from the quasar, because the redshift also includes information on its peculiar velocity.
If we did that, we would derive a proper distance of the order of $\sim$10 Mpc, implying a lack of correlation between the absorption system and the quasar host galaxy. Probably, all the gas explored in those works is contained in our first bin, within 5000 km s$^{-1}$ of z$\rm_{em}$. What is relevant is that we observe a dramatically different Mg{\sevensize II} covering fraction at any velocity separation that is reasonable to associate with gas within the host galaxy.

\subsection{Importance of barycentre determination}
\label{barycentre}

In the previous sections, we have considered the position of the deepest component of each C{\sevensize IV} system as the zero velocity point around which we select the regions to produce the stacked spectrum (as described in Section~\ref{stacking}). Such barycentre selection is justified by the fact that low-ionization ions in our sample generally appear to be well-aligned with the deepest component of C{\sevensize IV} systems that exhibit a complex velocity-component structure. Now, we explore how the stacking is affected by this choice (see also \citealp{Ellison2000}).

The deepest component of the C{\sevensize IV} systems in our sample not always coincide with the centre of their velocity profiles.
The most affected systems are those for which it is not possible to discern the individual narrow components within the adopted clustering velocity (see Section~\ref{civ_sample}). They are usually the strongest systems in a given velocity bin and dominate the final composite spectrum. As a consequence, the stacked absorption profiles appear slightly asymmetric (see Fig.~\ref{fig:0-5}). In contrast, this trend does not affect the weak systems, which we are able to consider separately from each other.

We now calculate the barycentre position of each C{\sevensize IV} system, weighting the wavelengths with the optical depth of the line profile. Fig.~\ref{fig:weighted} in the Appendix shows the results. This technique produces more symmetric absorption troughs for the intermediate- and high-ionization ions. However, their profiles are broader and less representative of the width of direct individual detections. In addition, this method spreads out the signal of the low-ionization ions, making their signatures less evident and in some cases off-centered. This latter effect happens because most of the low-ionization ions are associated with DLA systems, and their absorption profile is often asymmetric. 
In conclusion, selecting the barycentre of the regions to be stacked according to the minimum of each C{\sevensize IV} system, we produce much more coherent absorption signatures.

We also note that, while the N{\sevensize V} stacked spectrum matches the kinematic profile of the corresponding C{\sevensize IV}, the O{\sevensize VI} absorption trough appears more irregular and broad (see Fig.~\ref{fig:0-5}). N{\sevensize V} absorptions arise predominately at the same velocity as C{\sevensize IV}, while O{\sevensize VI} is often significantly shifted to positive velocities, at both high and low redshift (e.g. \citealp{FR09, Lopez2007, Carswell2002, Tripp2008} ).
The usual interpretation of this often observed velocity shift between O{\sevensize VI} and C{\sevensize IV} (e.g. \citealp{Reimers2001, Carswell2002, Simcoe2002}) is that the lines do not arise from the same volume.
However, the origin of the shift remains uncertain.

In summary, we find that the N{\sevensize V} features in our sample are well-aligned with C{\sevensize IV}, implying that both species arise from the same gas phase, whereas, velocity offsets between C{\sevensize IV} and O{\sevensize VI} centroids may contribute to producing a broader stacked profile compared to other species. 
We will discuss the possible origin of N{\sevensize V} and O{\sevensize VI} more extensively in Section~\ref{nature}.

\section{DISCUSSION}
\label{discussion}

In this article, we stack 100 quasar spectra in the rest-frame of C{\sevensize IV} absorbers detected in the dataset of XQ-100 Legacy Survey, to examine the typical strength of N{\sevensize V} and other lines in the vicinity of quasars compared with intervening gas. The combination of moderate resolution, high S/N and wide wavelength coverage of our survey have allowed us to build composite spectra at redshifts  2.55-4.73 and at the same time explore a wide range of strong and weak NALs.

From now on, we refer to the top panels of Fig.~\ref{fig:EW_ions} (panels a and e) to describe the associated region for strong and weak systems, respectively. On the other hand, the bottom panels (panels d and h), characterize intervening gas. Fig.~\ref{fig:EW_ions}(d) and (h) correspond to a velocity offset from z$\rm_{em}$ (10,000 < v$\rm_{abs}$ < 35,000 km s$^{-1}$) that includes both the smaller bins explored in the two upper panels and is large enough to allow the investigated ions to produce measurable absorptions. Besides this, we are confident that in Fig.~\ref{fig:EW_ions}(d) and (h) only intervening systems are selected, because this velocity bin does not encompass the excess of NALs within 10,000 km s$^{-1}$ of z$\rm_{em}$ found in P16 (see Fig.~\ref{fig:EW_5-10} for more details). DLAs have been excluded from the stacking, since they represent a special class of systems. 

AALs are characterized by higher ionization states than intervening absorbers. Indeed, we can see in Fig.~\ref{fig:EW_ions} that N{\sevensize V} exhibits a strong absorption signal only within 5000 km s$^{-1}$ of z$\rm_{em}$. This absorption trough is much stronger if C{\sevensize IV} systems with N(C{\sevensize IV}) > 10$^{14}$ cm$^{-2}$ are used to build the composite spectrum. We draw attention to the fact that O{\sevensize VI} is detected in the proximity of the quasar, as well as in the intervening gas. This is not surprising, since O{\sevensize VI} absorption is observed in a wide range of astrophysical environments (see \citealp{Fox2011} for a review). However, the O{\sevensize VI}/C{\sevensize IV} ratio is higher in AALs. Fig.~\ref{fig:EW_ions} also shows that Si{\sevensize IV} is detected both in the proximity of the quasar and in intervening gas. Nevertheless, it is much weaker close to the quasar, where the gas is more easily ionized, and it is thus reasonable to expect that a smaller fraction of silicon is in the form of Si{\sevensize IV}. We note that Si{\sevensize III} signal increases moving farther from the quasar and that the Si{\sevensize IV}/Si{\sevensize III} and Si{\sevensize IV}/C{\sevensize IV} ratios suggest lower ionization in intervening systems with respect to associated ones. This is true for the composite spectra obtained selecting both weak and strong C{\sevensize IV} systems. AALs are also characterized by the absence of the low-ionization ions. They are found, instead, at large velocity separation from the z$\rm_{em}$, but only associated with the strongest C{\sevensize IV} systems in the sample (i.e. N(C{\sevensize IV}) > 10$^{14}$ cm$^{-2}$). Indeed, low-ionisation metal transitions, such as Mg{\sevensize II}, C{\sevensize II} and Si{\sevensize II}, are mostly found at N(H{\sevensize I}) $\geq$ 10$^{16}$ cm$^{-2}$ (e.g. \citealp{Farina2014}).

\begin{figure}
 \includegraphics[width=\columnwidth]{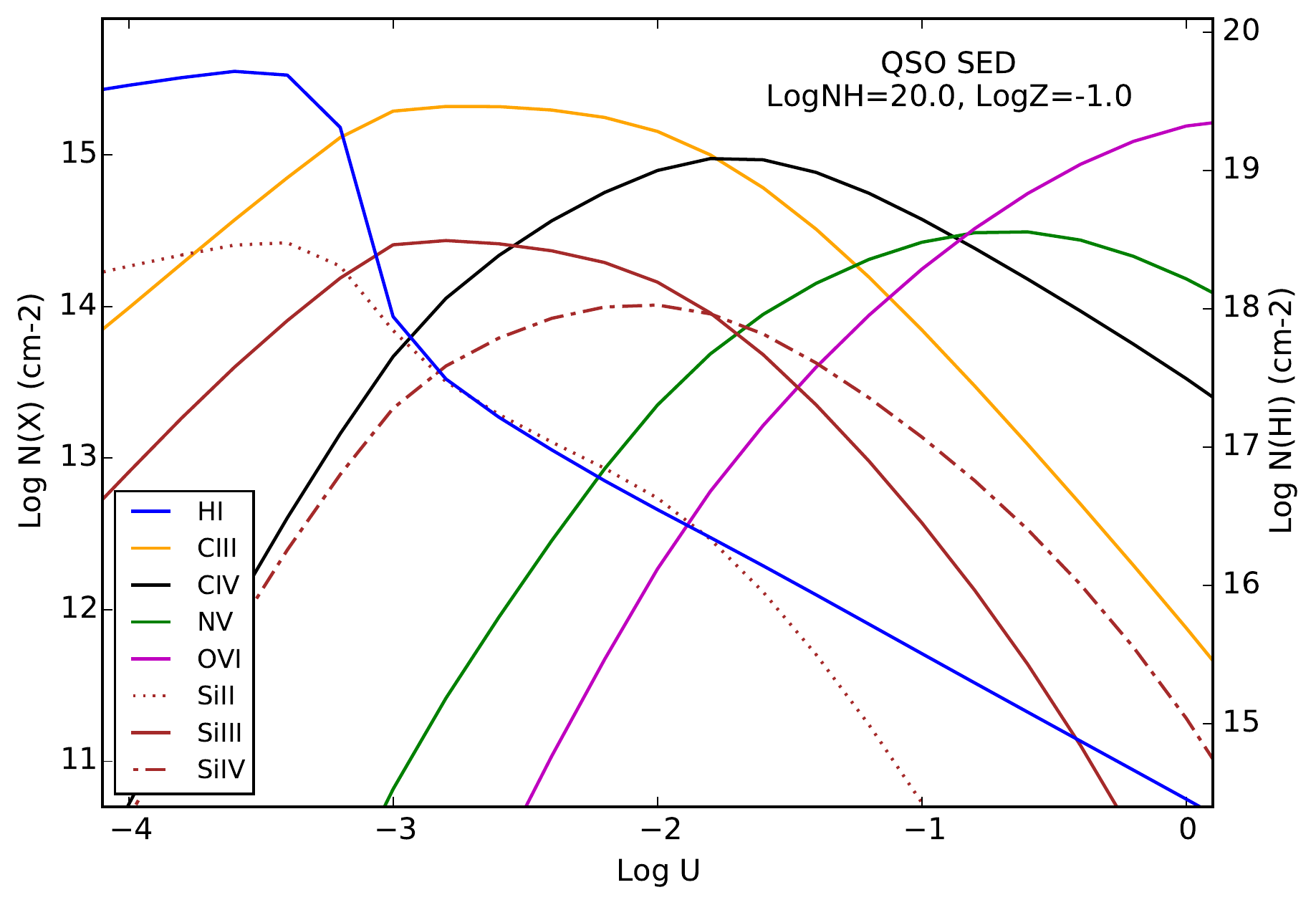}
 \includegraphics[width=\columnwidth]{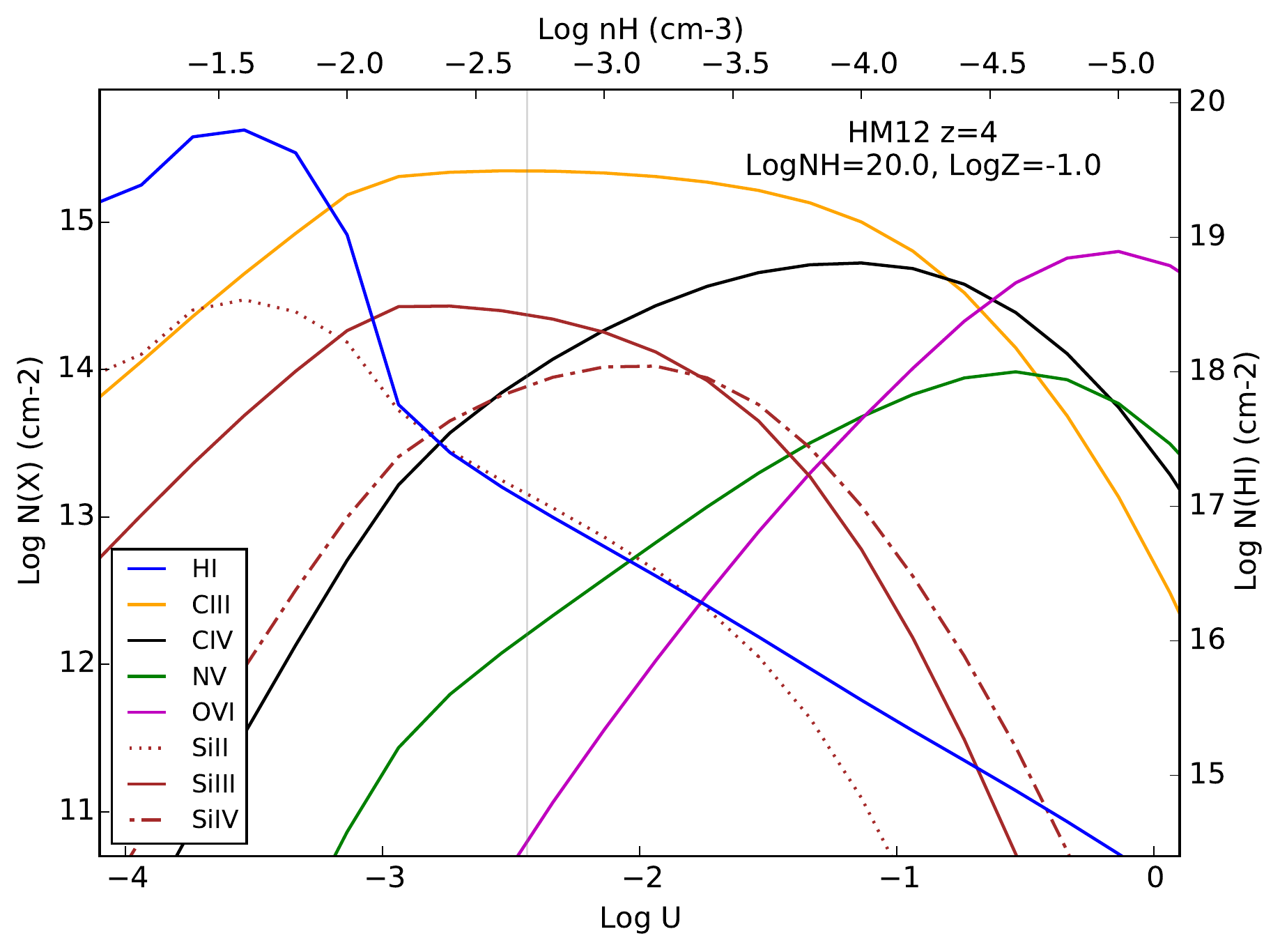}
\includegraphics[width=\columnwidth]{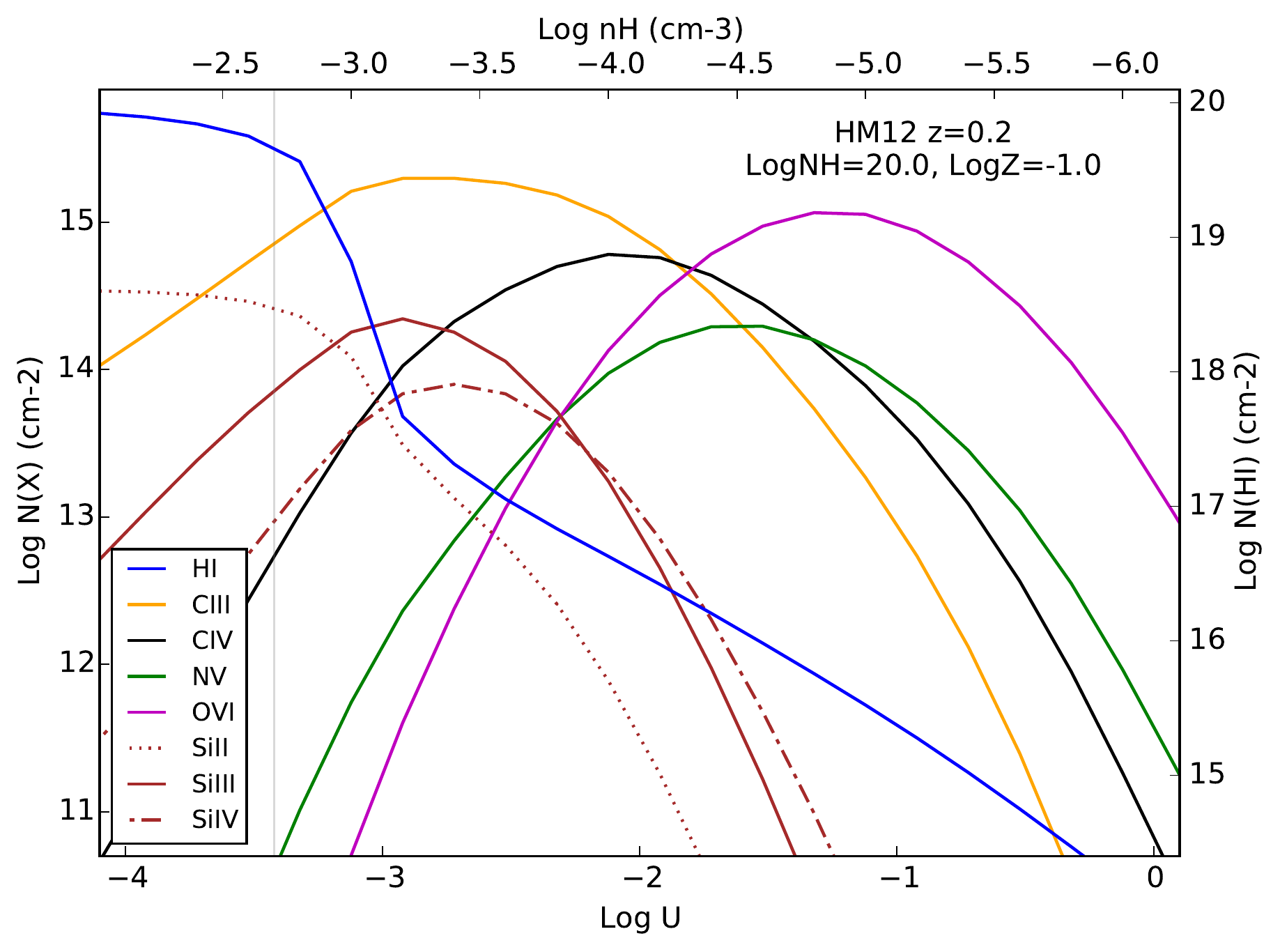}

 \caption{Cloudy-predicted column densities as a function of the ionization parameter U, for a cloud of total column density Log (N$\rm_H$) = 20 cm$^{-2}$. The metallicity of the cloud is Log Z = $-$1.0 with solar relative abundances. The N(H{\sevensize I}) is shown by the bold blue curve and axis labels on the right. Top panel: the gas is assumed to be photoionized by a quasar Spectral Energy Distribution. Middle and bottom panels: the gas is assumed to be photoionized by the UV background radiation \citep{Haardt2012} at z = 4 (middle panel) and z = 0.2 (bottom panel).}
 \label{fig:QSO_SED}
\end{figure}

\begin{figure*}
 \includegraphics[width=\columnwidth]{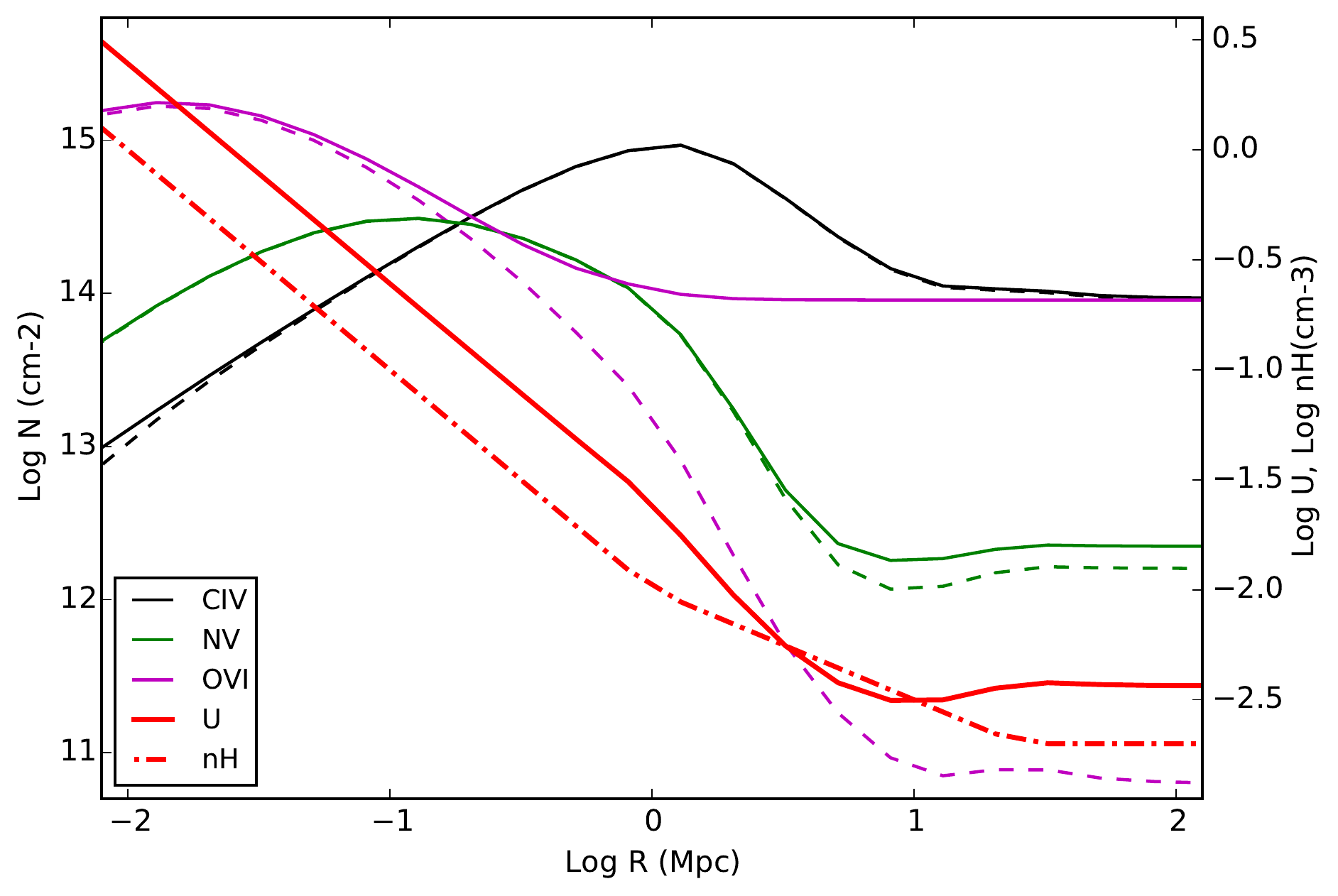}
 \includegraphics[width=\columnwidth]{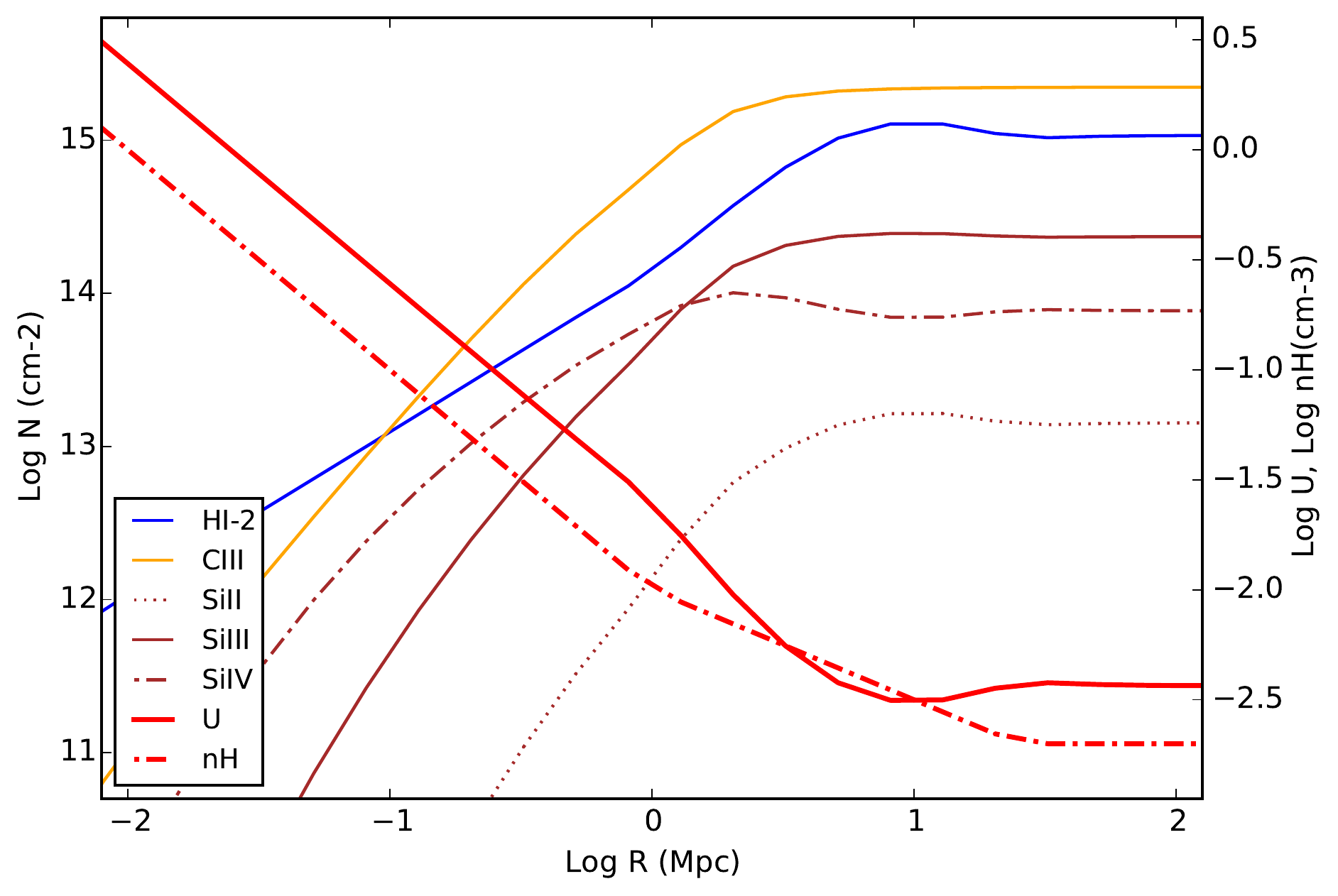}

 \caption{Cloudy-predicted column densities as a function of the distance, R, from a quasar at z = 4.The metallicity of the cloud is Log Z = $-$1.0 with solar relative abundances and the total column density is Log (N$\rm_H$) = 20. The gas is modelled as a uniform slab in thermal and ionization equilibrium and is assumed to be photoionized by the radiation field of a quasar with bolometric luminosity L = 10$^{47}$ erg s$^{-1}$ and the UV background radiation \citep{Haardt2012}. We also add a thermal gas component with Log T = 5.5. The solid curves show the sum of the two contributions, while the dashed ones show just the photoionized component. The ions in the right panel are not affected by the thermal component. The H{\sevensize I} column density, labeled \enquote{H1-2}, is plotted as Log N(H{\sevensize I}) $-$ 2.0 to fit on the plot. The densities of the cloud and the U parameters at each R are shown by the dot-dashed and bold red curves, respectively, and axis labels on the right. Left: C{\sevensize IV}, N{\sevensize V} and O{\sevensize VI} column densities. Right: low-ions column densities.}
 \label{fig:cloudy_R}
\end{figure*}

Fig.~\ref{fig:EW_ions} clearly shows that the gas in the proximity of the quasar exhibits a considerably different ionization state with respect to the intervening gas. In order to illustrate how the absorption-line strengths of the explored ions vary as a function of different physical conditions, we use the Cloudy software package (v17.00; \citealp{Ferland2017}) to run ionization models.

A photoionized plasma is characterized by the ionization parameter, U = Q$\rm_H$/4$\pi$R$^2$n$\rm_H$c, where Q$\rm_H$ is the source emission rate of hydrogen ionizing photons, R is the distance from the central source, n$\rm_H$ is the hydrogen density and c is the speed of light. Top panel of Fig.~\ref{fig:QSO_SED} exhibits how the column densities of several ions within a cloud of total column density Log N(H) = 20 cm$^{-2}$, vary as a function of U. In this model, the gas is assumed to be photoionized only by the radiation field of the quasar. The adopted bolometric luminosity of the quasar is L = 10$^{47}$ erg s$^{-1}$, representative of the XQ-100 sample (see P16). We consider a metallicity Log Z = $-$1.0, with solar relative abundances. We note that, close to the quasar (high values of U), the intense and hard radiation field eliminates the low ions and allows nitrogen and oxygen to be highly ionized. Moving farther from the quasar, the decrease of U produces the drop of the N{\sevensize V} and O{\sevensize VI} absorption signal, and the appearance of the low-ions.

Fig.~\ref{fig:cloudy_R} illustrates this trend even better, showing how the column densities of the same ions vary as a function of the distance, R, from a quasar at z = 4 (properties of the cloud as in Fig.~\ref{fig:QSO_SED}). The gas is assumed to be photoionized by the radiation field of the quasar and the UV background \citep{Haardt2012}. We also add a thermal gas component with Log T = 5.5 to mimic the effect of collisionally ionized gas. The dashed curves show the photoionized component, while the solid ones represent the sum of the two contributions. 

Let us focus on the photoionized component first. We can see in the left panel of Fig.~\ref{fig:cloudy_R} that N{\sevensize V} and O{\sevensize VI} (and slightly C{\sevensize IV}) drop with distance from the quasar, because the gas is subjected to lower values of U (it corresponds to moving to the left in the top panel of Fig.~\ref{fig:QSO_SED}). Those ions signal would continue to decrease, except that at large R they hit a floor determined by the UV background. Fig.~\ref{fig:spectra} in the Appendix shows that the UV background at z = 4 starts to dominate the ionizing continuum (> 13.6 eV) at R $\gtrsim$ 3 Mpc, but its spectral energy distribution (SED) is strongly shaped by deep edges due to H{\sevensize I} and He{\sevensize II} ionization. In particular, the C{\sevensize III} edge is at lower energies than the He{\sevensize II} one, so the UV background radiation at z = 4 can produce C{\sevensize IV}, but it cannot produce much O{\sevensize VI} and even less N{\sevensize V}. The right panel of Fig.~\ref{fig:cloudy_R} shows how, moving away from the quasar, the decreases of U allows low-ions absorption signal to rise. 

The trends illustrated in Fig.~\ref{fig:cloudy_R} give a natural explanation for the increase in Si{\sevensize IV}/C{\sevensize IV} and Si{\sevensize III}/Si{\sevensize IV} ratios we see in Fig.~\ref{fig:EW_ions} moving towards gas at large separation from the quasar. The substantial difference in the N{\sevensize V}/C{\sevensize IV} ratio between AAL and intervening systems is also well described. However, if the drop of U were the only factor in play, we should detect a much weaker O{\sevensize VI} signal far from the z$\rm_{em}$ in our composite spectra. Indeed, the N{\sevensize V} ionization potential is in between those of C{\sevensize IV} and O{\sevensize VI} and we can see in Fig.~\ref{fig:EW_ions} that N{\sevensize V} signal drops moving away from the quasar, while the O{\sevensize VI}/C{\sevensize IV} ratio is not subject to a change consistent with that of N{\sevensize V}. This trend can have multiple interpretations. 

One possibility is that O{\sevensize VI} traces another phase of the absorbing gas, independently from the environment, and it has a different mechanism of production with respect to N{\sevensize V}, so the change of the ionization radiation at large distance from the quasar affects only the N{\sevensize V} signal. This is unlikely, since O{\sevensize VI} is often found to be photoionized (e.g. \citealp{Bergeron2002, Reimers2006, FR09}). It is also possible that O{\sevensize VI} is photoionized in the proximity of the quasar and has a different channel of production in other environments.

Fig.~\ref{fig:cloudy_R} clearly illustrates that a thermal gas component with Log T = 5.5 increases N(O{\sevensize VI}) significantly at large R (magenta curves), but it contributes little to N(N{\sevensize V}) (green) and negligibly to N(C{\sevensize IV}) (black). This model suggests that, moving farther from the quasar, the change in the U parameter causes the drop in both N{\sevensize V} and O{\sevensize VI}. The fact that we see an O{\sevensize VI} signal in our composite spectra even at large velocity offsets from z$\rm_{em}$ can be explained by an additional thermal gas component.

Right panel of Fig.~\ref{fig:EW_ions} shows that the lower N(C{\sevensize IV}) systems have much higher C{\sevensize III}/C{\sevensize IV} than the high N(C{\sevensize IV}). Therefore, selecting weak C{\sevensize IV} absorbers we do not simply find lower total column densities. We also find systems with lower ionization. However, this trend cannot be explained only by changing U, because the strong C{\sevensize III} and larger C{\sevensize III}/C{\sevensize IV} ratios are not (clearly) accompanied by stronger absorption in other lower ions. All of these absorbers can have a range of n$\rm_H$, which means a range in U values. So, from a distribution of absorbers, systems with weak C{\sevensize IV} NALs and small C{\sevensize IV}/C{\sevensize III} ratios might have relatively less of the low n$\rm_H$, high U gas (not simply lower U overall).

\subsection{NV and OVI as diagnostics of ionization mechanisms}
\label{nature}

Conclusions regarding the origin and fate of CGM are linked to the assumptions one makes about the physical processes that determine its ionization state. 
Successful models of the CGM must account for the full suite of observations available and combine together the two dominant gas phases, typically described as a cool $\sim$ 10$^4$ K and a warm $\sim$ 3 $\times$ 10$^5$ K phase.
In this context, the presence or absence of N{\sevensize V} and O{\sevensize VI} in quasar absorption systems may provide crucial information for understanding the physical conditions of the absorbing gas. N{\sevensize V} and O{\sevensize VI} require 77.5 (97.9) and 113.9 (138.1) eV to be created via photoionization (to be destroyed), respectively, and their abundance peaks at 2 $\times$ 10$^5$ and 3 $\times$ 10$^5$ K for collisionally ionized gas \citep{Gnat2007}. 

\citet{Lu1993} performed a search for N{\sevensize V} and O{\sevensize VI} absorption based on a composite spectrum method. To this aim they exploited a total of 73 intervening C{\sevensize IV} systems (with mean redshift $\langle$z$\rangle$ = 2.76) identified in 36 spectra from the survey by \citet{Sargent1989}.  The results of this study may apply to relatively strong systems, since only C{\sevensize IV} absorbers with rest-frame equivalent width $\geq$ 0.3 \AA\, can be detected in those spectra. This work provided the first firm evidence for the presence of the O{\sevensize VI} absorption in intervening quasar metal line systems, while N{\sevensize V} doublet was not detected. This agrees with what we find in our composite spectrum obtained selecting only strong C{\sevensize IV} absorbers. They estimated the ratio N(O{\sevensize VI})/N(N{\sevensize V}) to be $\geq$ 4.4, which, for collisionally ionized gas in thermal equilibrium with a solar ratio of oxygen to nitrogen, implies a gas temperature T $\geq$ 2.5 $\times$ 10$^5$ K.

More recently, many detailed studies have been carried out on an absorber-by-absorber basis in quasar spectra that trace low-density foreground gas in the IGM and galaxy halos. 
These studies have shown that the ionization mechanisms of high-ionization species like O{\sevensize VI} and N{\sevensize V} are complex and vary over a wide range of environments.
Line diagnostics from low, intermediate, and high ions sometimes support a similar, photoionized origin for O{\sevensize VI}, N{\sevensize V}, and low-ionization state gas (e.g. \citealp{Tripp2008, Muzahid2015}) and sometimes require O{\sevensize VI} to be ionized by collisions of electrons with ions in a $\sim$ 3 $\times$ 10$^5$ K plasma (e.g. \citealp{Tumlinson2005, Fox2009, Wakker2012, Meiring2013}).

Recently, the work by \citet{Werk2016} reported on quasars probing sight lines near 24 star-forming galaxies with L $\approx$ L$^*$ at z = 0.2 from the COS-Halos survey \citep{Werk2013, Tumlinson2013}, which exhibit positive detections of O{\sevensize VI} in their inner CGM. Of the galaxies analysed, only three have positive detections of N{\sevensize V}. They also argue about the possibility that N{\sevensize V} does not trace the same gas phase as O{\sevensize VI}, since they show different velocity profiles. However, a N{\sevensize V} upper limit can impose meaningful constraints in both photoionized and collisionally ionized gas models. For example, the upper limits on Log N(N{\sevensize V})/N(O{\sevensize VI}) $\sim$ $-$1 strongly rule out most photoionization models, both in and out of equilibrium. 
\citet{Bordoloi2017} claim that most of the observed COS-Halos O{\sevensize VI} absorption-line systems can be explained as collisionally ionized gas radiatively cooling behind a shock or other type of radiatively cooling flow. Their model predicts the typical N{\sevensize V} column densities to be an order of magnitude smaller than the O{\sevensize VI} ones at similar cooling velocities. Therefore, most of the COS-Halos N{\sevensize V} systems associated with O{\sevensize VI} absorption would reside below the detection threshold, consistent with N{\sevensize V} non-detection for the majority of the COS-Halos O{\sevensize VI}. In the context of a multiphase CGM, however, this scenario does not take into account the constraints on the N{\sevensize V} signal coming from photoionization models based on the low and intermediate ions.

Fig.~\ref{fig:QSO_SED} shows the Cloudy-predicted column densities of several ions as a function of the ionization parameter U, for a cloud of total column density Log (N$\rm_H$) = 20 cm$^{-2}$, at z = 4 (middle panel) and z = 0.2 (bottom panel). We model a cloud with the same parameters as in Fig.~\ref{fig:cloudy_R}, to investigate how the picture described at large separation from the quasar changes with redshift. The vertical grey line in the middle panel marks U and n$\rm_H$ reached at large distance in Fig.~\ref{fig:cloudy_R}. We note that, considering the same U parameter, the \citet{Haardt2012} spectrum produces more N{\sevensize V} and O{\sevensize VI} at low z. Fig.~\ref{fig:spectra} in the Appendix shows that the UV background flux at energies from $\sim$13.6 eV to $\sim$45 eV is weaker at z = 0.2 by 0.98 dex compared with z = 4. However, for any given density, it produces $\sim$10 times lower U at z = 0.2 than at z = 4, but more N{\sevensize V} and O{\sevensize VI}, due to the harder continuum shape around the energies that create those ions.
Single-phase photoionization models fail in explaining the range of N(N{\sevensize V})/N(O{\sevensize VI}) reported by studies at low z (e.g. COS-Halos survey). As discussed in \citet{Werk2016}, the observed Log N(N{\sevensize V})/N(O{\sevensize VI}) $\sim$ $-$1 would require very low gas density for the CGM with U $\geq$ $-$1 (see the bottom panel of Fig.~\ref{fig:QSO_SED}). 
However, depending on the physical condition of the cloud, the N{\sevensize V} signal produced by photoionization can be an important contribution to the total strength of the absorbing line. Therefore, before using the N{\sevensize V} non-detection to derive a limit on the gas temperature, one can exploit the constraints from low and intermediate ions to predict the N{\sevensize V} absorption produced by the UV background radiation.

The details of the mechanisms that ionize the absorbing gas can rapidly get complex if non-equilibrium processes are occurring and/or if there is a combination of photoionization and collisional ionization (\citealp{Savage2014}, for a recent review).
Among the models proposed to reproduce the intermediate- and high-ion states (e.g. C{\sevensize IV}, N{\sevensize V}, O{\sevensize VI}) there are (1) time-variable radiation fields from AGN coupled with non-equilibrium ionization effects \citep{Segers2017, Oppenheimer2017}, (2) radiative cooling flows \citep{Bordoloi2017, McQuinn2018}, (3) formation in fast shocks \citep{Gnat2009} or (4) formation within thermally conductive interfaces \citep{Gnat2010, Armillotta2017}.

We have shown that it is possible that N{\sevensize V} is photoionized in environments where O{\sevensize VI} is mostly collisionally ionized, but being weak it is hard to detect. We have also demonstrated that the observed differences in the N{\sevensize V}/C{\sevensize IV} ratio between AALs and intervening systems can be explained naturally by ionization effects. However, another factor that can contribute in making the intervening N{\sevensize V} signal weaker than already expected is that these absorbers can trace gas in regions with lower metallicity or sub-solar N/$\alpha$ relative abundance. 
N{\sevensize V} absorptions are largely identified in proximity to bright quasars along the line of sight (e.g. \citealp{Misawa2007, Nestor2008, Ganguly2013}, P16, \citealp{Chen2017}). Quasars are generally metal-rich and about solar [N/H] has been measured in associated systems \citep{Petitjean1994, Dietrich2003, Dodorico2004, Ganguly2006, Schaye2007}, making N{\sevensize V} easily detectable in proximate absorbers. Nitrogen is rarely detected in optically thin (i.e. N(H{\sevensize I}) $\lesssim$ 10$^{17}$ cm$^{-2}$) intergalactic absorption systems. In the IGM, the metallicity and therefore the nitrogen content is lower, thus N{\sevensize V} features are expected to be weak \citep{FR09}.

\citet{FR09} find that the [N/$\alpha$] abundance in intervening systems is clearly sub-solar, with a median [N/$\alpha$] = $-$0.58 and values down to $\sim$ $-$1.8. Observations have shown that the N/O ratio can diminish moving outwards of disc galaxies as a function of the galactocentric distance and stellar mass (e.g. \citealp{Belfiore2017}). This result is supported by chemical evolution models included within cosmological hydrodynamical simulations  \citep{Vincenzo2018a,Vincenzo2018b}.   
\enquote{Down the barrel} observations give us access to gas originated at small distances from the centre of the quasar host galaxy (e.g. \citealp{Wu2010}). In contrast, the absorption systems identified in works based on projected pairs usually have impact parameters from tens to hundreds of kpc from the z$\rm_{em}$ of the background target, so they trace gas in the outskirts of galaxies possibly characterized by a lower N/O ratio.

\section{CONCLUSIONS}
\label{conclusions}
In this work, we present new measurements of metal NALs made using composite quasar spectra. Our sample includes 100 individual lines of sight from the XQ-100 Legacy survey, at emission redshift z$\rm_{em}$ = 3.5-4.72. To build the stacking spectra, we start from a large number ($\sim$ 1000) of C{\sevensize IV} absorption systems identified in P16, covering the redshift range 2.55 < z$\rm_{abs}$ < 4.73. The main goal of our analysis is to investigate the N{\sevensize V} absorption signal at large velocity separations from z$\rm_{em}$. This study  complement our previous work on N{\sevensize V} and test the robustness of the use of N{\sevensize V} as a criterion to select intrinsic NALs. We also characterize the ionization state of the gas, both near and at great distance from the quasar. Our primary results are as follows.

(1) We show the absence of a statistically significant intervening N{\sevensize V} absorption 
signal along the line of sight of background quasars. Indeed, the N{\sevensize V} exhibits a strong absorption trough only within 5000 km s$^{-1}$ of z$\rm_{em}$. This feature is much stronger ($\sim$ 15$\sigma$ confidence level) if C{\sevensize IV}
 systems with N(C{\sevensize IV}) > 10$^{14}$ cm$^{-2}$ are considered to build the composite spectrum. This result supports our previous claim (see P16) that N{\sevensize V} is an excellent statistical tool for identifying intrinsic systems.
 
 (2) We use photoionization models to show that, in a scenario where associated systems are photoionized by a quasar, the ionization parameter is expected to drop dramatically with distance from the continuum source. This is the main reason for intervening N{\sevensize V} systems being weaker than associated ones. Another possible contribution to this trend is a lower metallicity or N/O abundance.
 
( 3) The gas close to the continuum source (v$\rm_{abs}$ < 5000 km s$^{-1}$) exhibits a different ionization state with respect to the intervening one. Moving farther from the quasar, we can appreciate the appearance of the low-ions associated with the strongest systems in the sample (N(C{\sevensize IV}) > 10$^{14}$ cm$^{-2}$) and the drop in the N{\sevensize V} absorption signal. In contrast, the O{\sevensize VI} NAL is detected even at large velocity shifts from z$\rm_{em}$. We also note that O{\sevensize VI} and C{\sevensize IV} do not exhibit a change in their absorption signal in line with the decrease of the N{\sevensize V} signal. We run photoionization models that describe these trends well (see Fig.\ref{fig:cloudy_R}). We use these models to illustrate that a drop in the U parameter cannot be the only element to explain the data and that an additional thermal component (T $\sim$ 3 $\times$ 10$^5$ K) can produce the O{\sevensize VI} absorption we detect at large distance from the quasar. We show how it is possible that N{\sevensize V} is photoionized in environments where O{\sevensize VI} is mostly collisionally ionized, but, being weak, this can be challenging to detect. 

(4) We verify the deficiency of cool gas, as traced by low ions and in particular by Mg{\sevensize II}, in the proximity of the quasar. This finding is in agreement with the dearth of CII NALs (as shown in P16) and confirms that the gas in the proximity of the quasar along the line of sight has a different ionization state with respect to the gas in the transverse direction (e.g. \citealp{Prochaska2014, Farina2014}).

\section*{Acknowledgements}

The XQ-100 team is indebted to the European Southern Observatory (ESO) staff for support through the Large Program execution process. 





\bibliographystyle{mnras}
\nocite{*}
\bibliography{Perrotta_2018} 



\appendix
\section{}
\label{sec:abbreviations}

\begin{figure*}
 \includegraphics[width=\columnwidth]{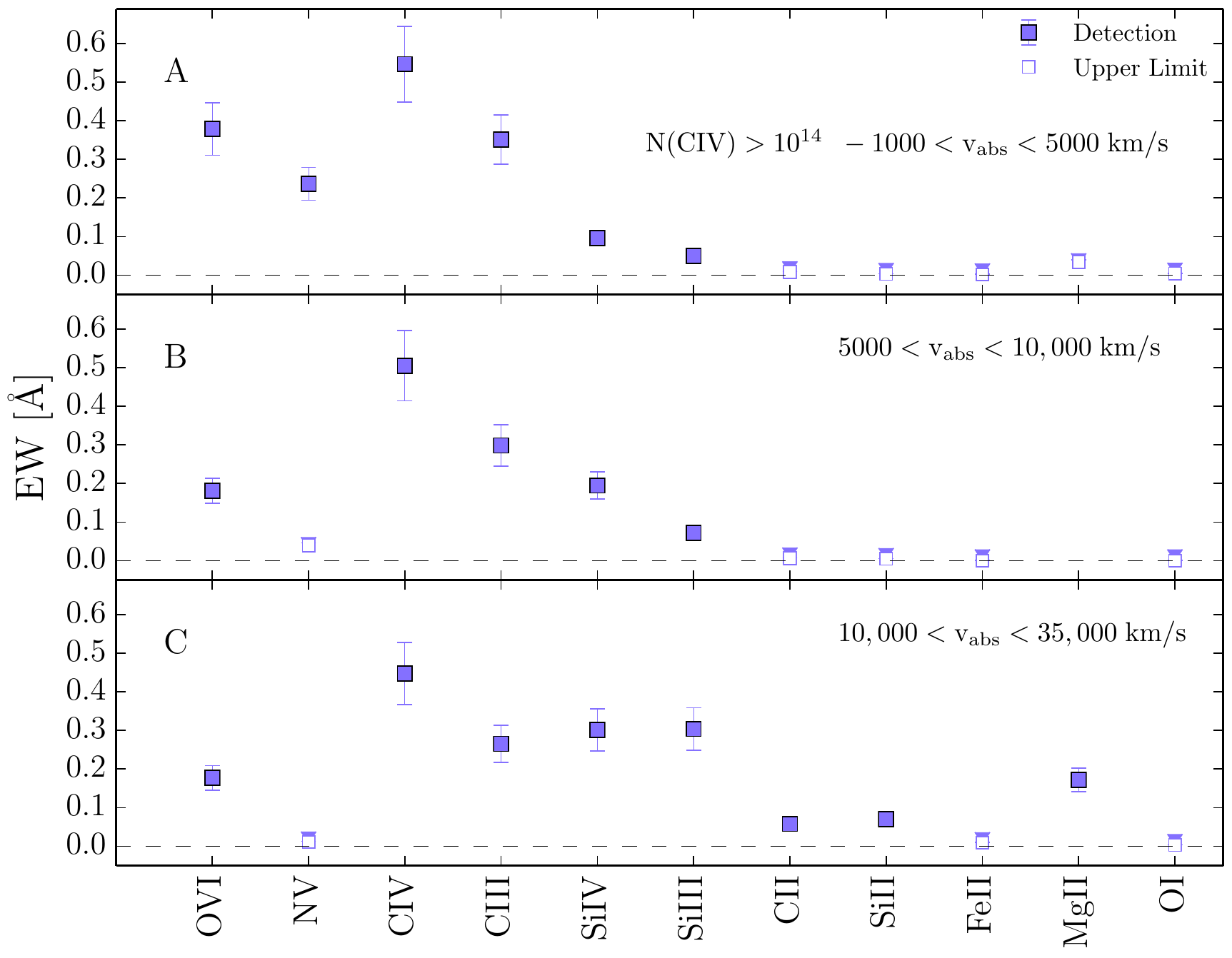}
 \includegraphics[width=\columnwidth]{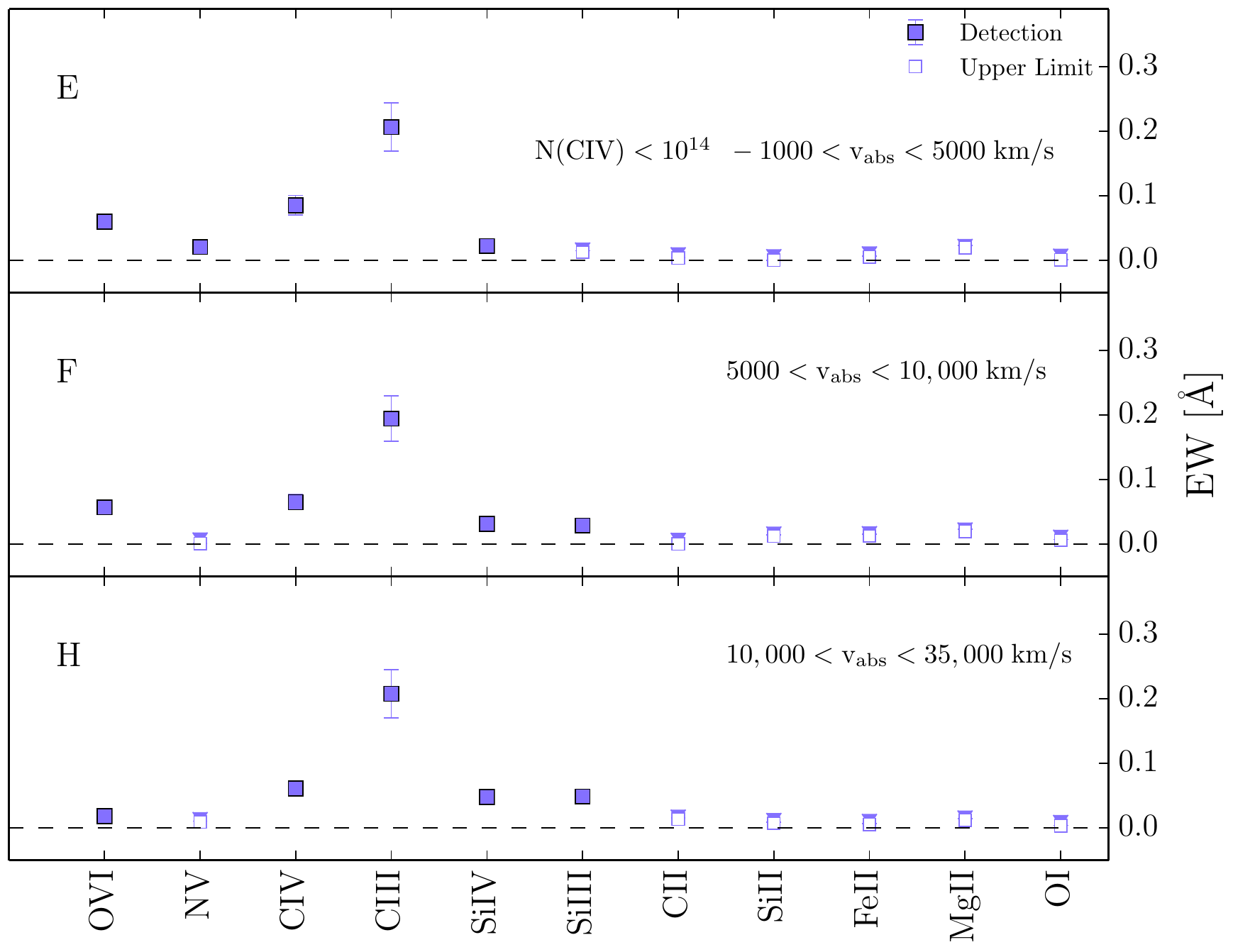}

 \caption{ Left: The equivalent widths of metal species measured in order of decreasing ionization potential for composite spectra based on the C{\sevensize IV} systems with N(C{\sevensize IV})$\geq$ 10$^{14}$ cm$^{-2}$. Each panel displays the velocity offset from z$\rm_{em}$. C{\sevensize IV} DLAs have been excluded form the stacking. Right: same as for the left panels, but in this case N(C{\sevensize IV})< 10$^{14}$ cm$^{-2}$ is considered as threshold for the stacking. The EW refers only to the strongest member of the doublets. Error bars are 1$\sigma$ and are estimated by bootstrapping the data using 1000 realizations.}
 \label{fig:EW_5-10}
\end{figure*}

\begin{figure*}
 \includegraphics[width=1.1\columnwidth]{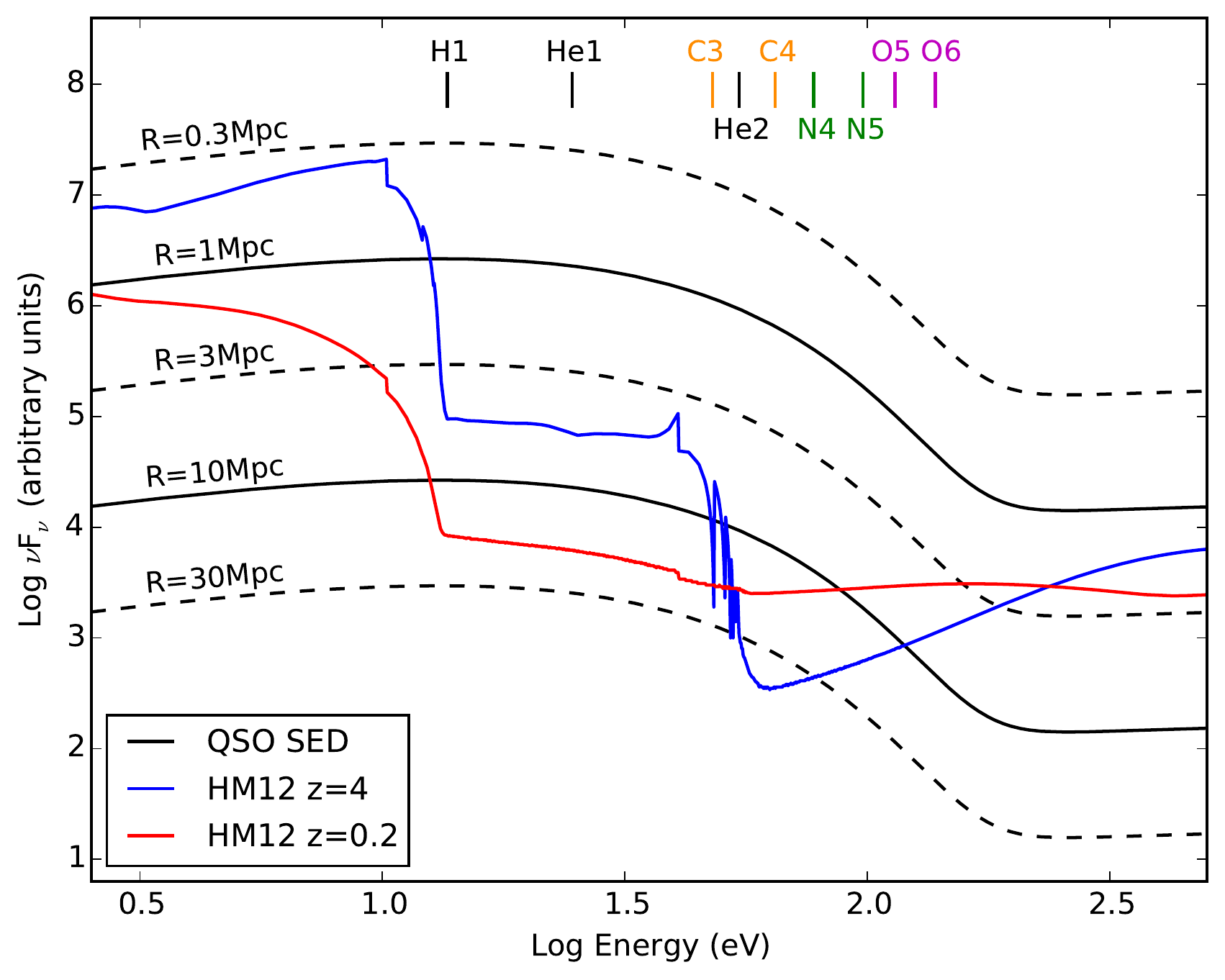}

 \caption{UV background \citep{Haardt2012} spectral energy distribution (SED) at z = 4 and z = 0.2, and quasar SED scaled to different distances from the quasar. The tick marks along the top mark the ionization energies of some ions important for the discussion.}
 \label{fig:spectra}
\end{figure*}

The left panels of Fig.~\ref{fig:EW_5-10} show the rest-frame equivalent widths of metal species measured in the composite spectrum based on C{\sevensize IV} systems with N(C{\sevensize IV}) $\geq$ 10$^{14}$ cm$^{-2}$ and 5000 < v$\rm_{abs}$ < 10,000 km s$^{-1}$ (panel b), compared to results for AAL and intervening systems (panels a and c), already presented in Fig.~\ref{fig:EW_ions}. The right panels are the same as the left panels, but in this case N(C{\sevensize IV}) < 10$^{14}$ cm$^{-2}$ is considered as threshold for the stacking. The 5000 < v$\rm_{abs}$ < 10,000 km s$^{-1}$ is a particular velocity bin in our study. In P16 we find that the C{\sevensize IV} sample exhibits a statistically significant excess ($\sim$ 8$\sigma$) within 10,000 km s$^{-1}$ of z$\rm_{em}$ with respect to the random occurrence of NALs, which is especially evident for lines with EW < 0.2 \AA. These latter lines more or less correspond to the weak systems in the current study. The excess extends to larger velocities than the classical associated region, is not an effect of NAL redshift evolution and does not show a significant dependence from the quasar bolometric luminosity.

We note that the systems in the 5000 < v$\rm_{abs}$ < 10,000 km s$^{-1}$ bin exhibit some of the properties of the associated systems and at the same time they show some characteristics of the intervening ones. Let us focus on the strong systems first. Fig.~\ref{fig:EW_5-10}(b) shows that the C{\sevensize IV}/Si{\sevensize IV} and C{\sevensize IV}/Si{\sevensize III} ratios are close to those of the AALs (Fig.~\ref{fig:EW_5-10}a), but O{\sevensize VI}/C{\sevensize IV} and N{\sevensize V}/C{\sevensize IV} ratios are in line with those of the intervening gas (Fig.~\ref{fig:EW_5-10}c). In addition, we only measure upper limits for the low ions, in agreement with gas in the proximity of the quasar. If we consider weak systems, Fig.~\ref{fig:EW_5-10}(f) shows a smaller C{\sevensize IV}/Si{\sevensize III} ratio like intervening gas, but a larger value of the O{\sevensize VI} absorption like the AALs.

Fig.~\ref{fig:EW_5-10}(b) and (f) seem to represent a "transition" region between associated and intervening systems. The EWs shown in those bins are produced by averaging together absorbers physically connected with the quasar host galaxy and its halo and spurious intervening systems. This is the main reason why we excluded this bin from our stacking analysis.

\begin{figure*}
 \includegraphics[width=\textwidth]{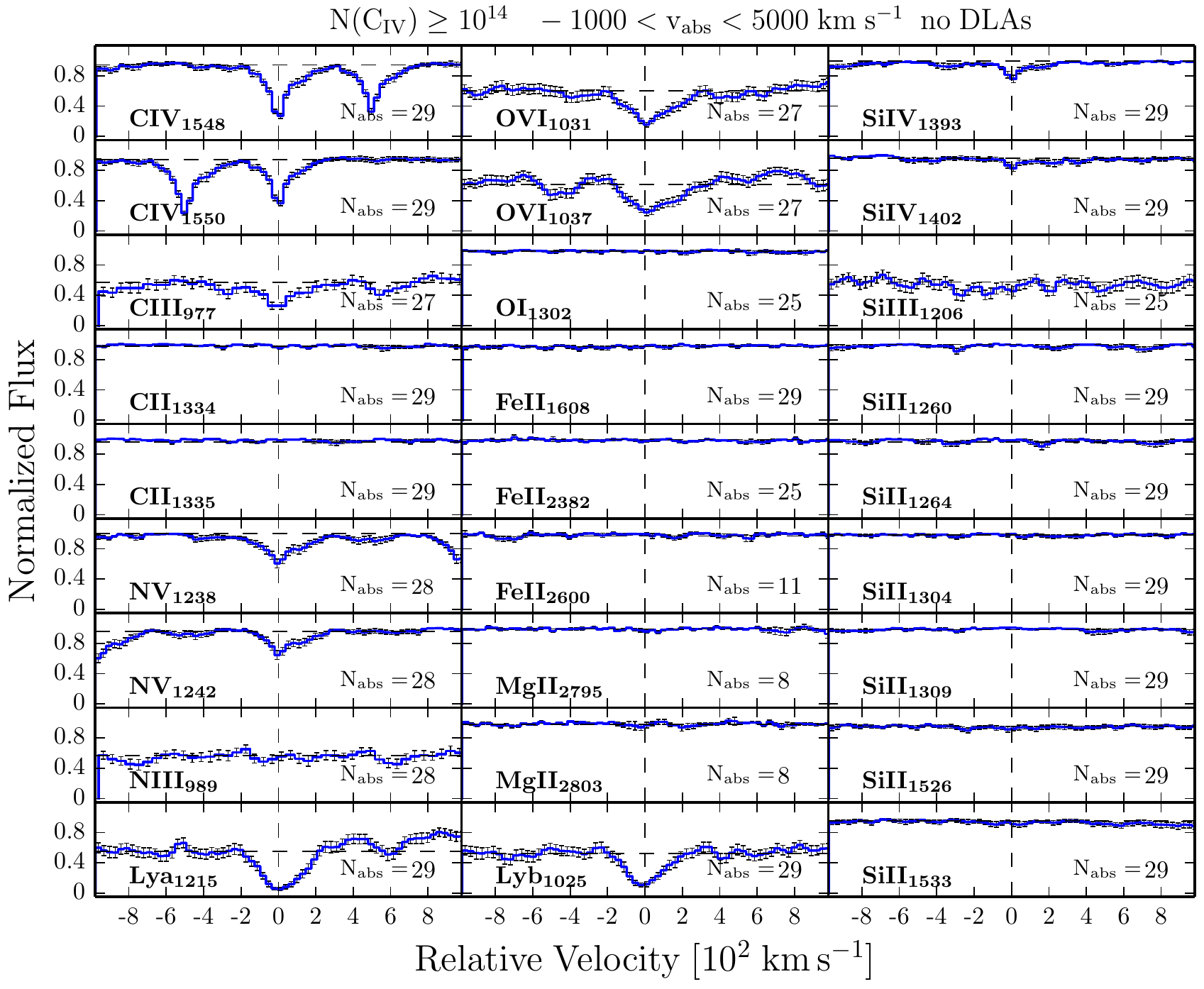}

 \caption{Composite spectra for the ions studied in this work, based on the C{\sevensize IV} systems with N(C{\sevensize IV})$\geq$ 10$^{14}$ cm$^{-2}$ and -1000 <v$\rm_{abs}$< 5000 km s$^{-1}$. C{\sevensize IV} DLAs are excluded from the stacking. Every panel reports the number of absorbers considered to build the stacked spectrum. The error bars are 1$\sigma$ and are estimated by bootstrapping the data using 1000 realizations.}
 \label{fig:0-5_nodla}
\end{figure*}

\begin{figure*}
 \includegraphics[width=\textwidth]{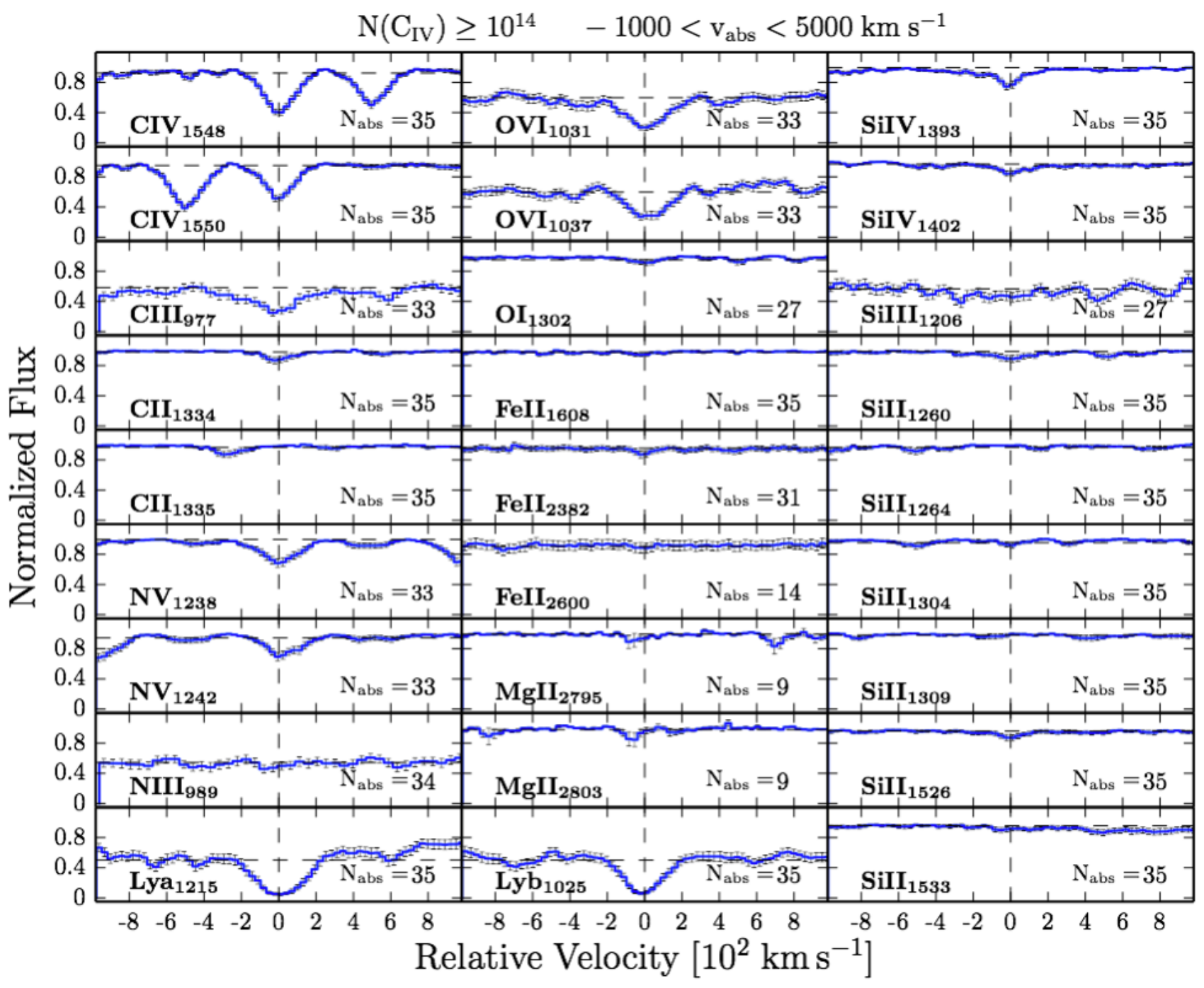}

 \caption{Composite spectra for the ions studied in this work, based on the C{\sevensize IV} systems with N(C{\sevensize IV})$\geq$ 10$^{14}$ cm$^{-2}$ and -1000 <v$\rm_{abs}$< 5000 km s$^{-1}$. The barycentre position of each C{\sevensize IV} system has been computed weighting the wavelengths with the optical depth of the line profile. Every panel reports the number of absorbers considered to build the stacked spectrum. The error bars are 1$\sigma$ and are estimated by bootstrapping the data using 1000 realizations.}
 \label{fig:weighted}
\end{figure*}

\begin{figure*}
 \includegraphics[width=\textwidth]{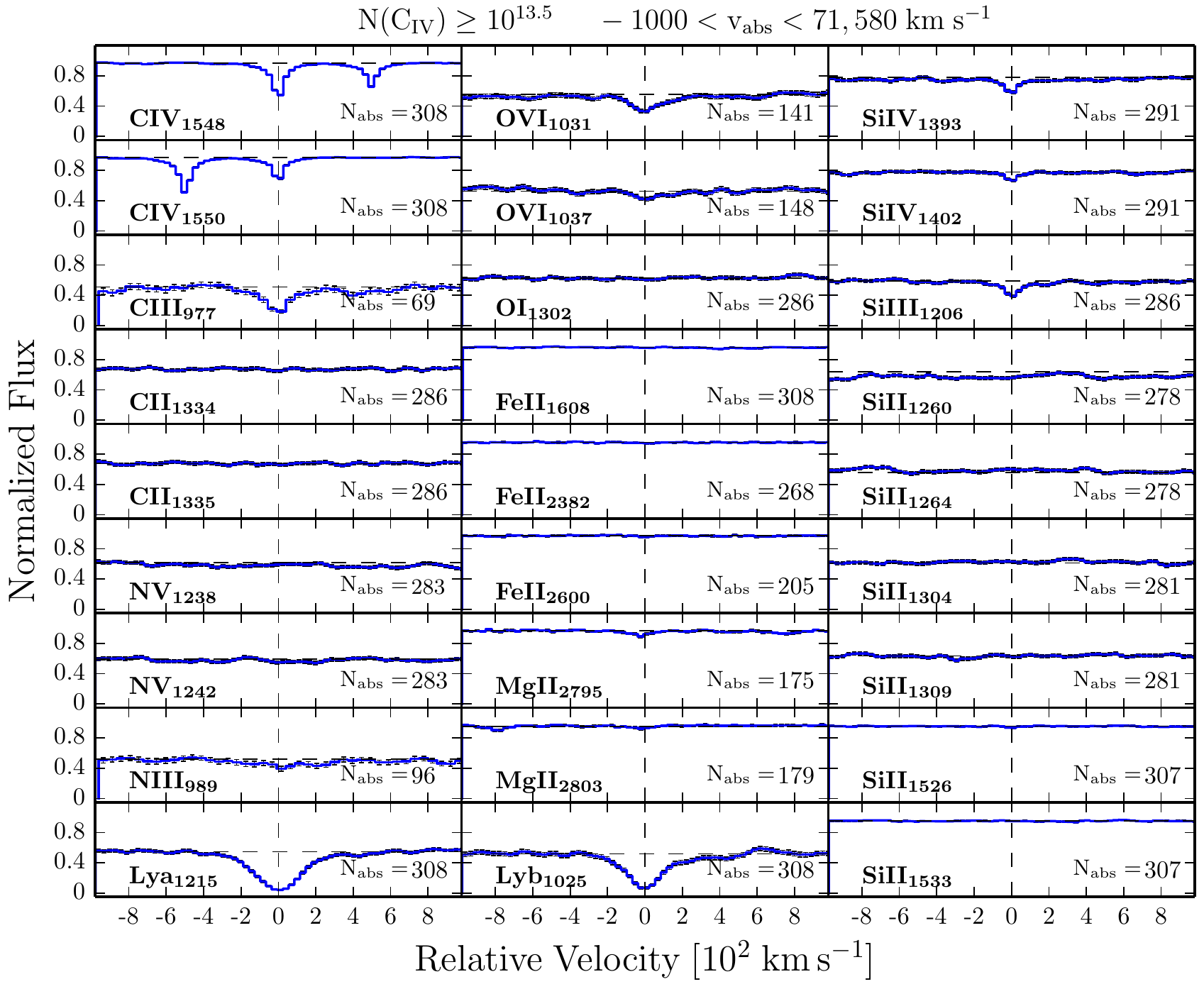}

 \caption{Composite spectra for the ions studied in this work, based on the C{\sevensize IV} systems with N(C{\sevensize IV})$\geq$ 10$^{13.5}$ cm$^{-2}$ and 15,000 < v$\rm_{abs}$< 71,580 km s$^{-1}$. C{\sevensize IV} DLAs are excluded by the stacking. Every panel reports the number of absorbers considered to build the stacked spectrum. The error bars are 1$\sigma$ and are estimated by bootstrapping the data using 1000 realizations. This example demonstrates that the stacking technique applied to the XQ-100 spectra can succeed in detecting absorption lines in the Ly$\alpha$ forest.}
 \label{fig:15-75_nodla}
\end{figure*}


\bsp	
\label{lastpage}
\end{document}